# Improving Biomedical Knowledge Graph Quality: A Community Approach

Katherina G Cortes [1], Shilpa Sundar[2], Sarah Gehrke[3], Keenan Manpearl [1], Junxia Lin [1], Daniel Robert Korn[3], Harry Caufield [4], Kevin Schaper [3], Justin Reese [4], Kushal Koirala [5], Lawrence E Hunter [6], E. Kathleen Carter [7], Marcello DeLuca [8], Arjun Krishnan [1], Chris Mungall [4], Melissa Haendel [3]

[1] Department of Biomedical Informatics, University of Colorado Anschutz Medical Campus
[2] Carolina Health Informatics Program, University of North Carolina at Chapel Hill
[3] TISLab, Department of Genetics, University of North Carolina at Chapel Hill
[4] Environmental Genomics and Systems Biology, Lawrence Berkeley National Laboratory
[5] Curriculum in Bioinformatics and Computational Biology, University of North Carolina at Chapel Hill
[6] Department of Pediatrics, University of Chicago
[7] Renaissance Computing Institute, University of North Carolina at Chapel Hill
[8] Eshelman School of Pharmacy, University of North Carolina at Chapel Hill

# Abstract

Biomedical knowledge graphs (KGs) are widely used across research and translational settings, yet their design decisions and implementation are often opaque. Unlike ontologies that more frequently adhere to established creation principles, biomedical KGs lack consistent practices for construction, documentation, and dissemination. To address this gap, we introduce a set of evaluation criteria grounded in widely accepted data standards and principles from related fields. We apply these criteria to 16 biomedical KGs, revealing that even those that appear to align with best practices often obscure essential information required for external reuse. Moreover, biomedical KGs, despite pursuing similar goals and ingesting the same sources in some cases, display substantial variation in models, source integration, and terminology for node types. Reaping the potential benefits of knowledge graphs for biomedical research while reducing duplicated effort requires community-wide adoption of shared criteria and maturation of standards such as Biolink and KGX. Such improvements in transparency and standardization are essential for creating long-term reusability, improving comparability across resources, providing a rigorous foundation for artificial intelligence models, and enhancing the overall utility of KGs within biomedicine.

# Background and Previous Work

Knowledge graphs (KGs) have gained popularity in recent years for their ability to connect different types and sources of data, offering a powerful computational framework to uncover novel biological hypotheses. Biomedical KG entities consist of disease, gene, mechanism, and other biological types. “Properties” linking entities include relationships such as ‘expressed in,’ ‘has phenotype’, or ‘interacts with.’ Biomedical KGs are typically constructed by aggregating and harmonizing existing curated knowledge bases and ontologies, supplemented by additional information that may come from text mining, data analyses, or case-level information. By ingesting and harmonizing many sources of knowledge, KGs are able to connect knowledge across biological domains leveraging existing research. For instance, the ROBOKOP (Reasoning Over Biomedical Objects linked in Knowledge Oriented Pathways)[1,2] KG platform enables researchers to query complex biomedical questions, such as “What environmental exposures are associated with asthma through intermediate biological processes?”, by integrating data from curated databases like CTD, DrugBank, and gene ontology (GO) annotations [3–5]. KGs have gained particular traction in biomedical research domains including drug repurposing, disease mechanism discovery, and rare disease illumination. Previous approaches have included using embeddings, vector representations of the nodes and edges which can then be used as input for artificial intelligence (AI) algorithms, of these KGs[6–9]. Other examples include the Monarch Initiative KG connects genotypic and phenotypic data across species to assist in the diagnosis of rare diseases, enabling clinicians to match patient symptoms with relevant genes even when traditional diagnostic methods fail [10,11]. Similarly, OpenTarget’s KG prioritizes drug targets by synthesizing genetic associations, expression data, and pathway information [12]. Many methods have been proposed to link knowledge graphs with large language models (LLMs) to reduce hallucinations and improve scientific utility[13–15]. As LLMs become more widely used in biomedicine, KGs can further support them through retrieval-augmented generation (KG-RAG), helping to normalize and ground generative outputs. Early efforts, such as pairing BioGPT and SciSpaCy[16,17] with structured KG-based retrieval, demonstrated how graphs could enhance scientific question answering and mechanistic reasoning. Today, more advanced approaches, such as DRAGON-AI, OntoGPT, AMG-RAG, DR.KNOWS, and KGARevion [18–22], illustrate a new generation of agentic KG-LLM integrations. These systems extend beyond retrieval to provide reliability, provenance, reasoning transparency, and medical context. Graph-based approaches offer scalable, efficient, and cost-effective ways to generate testable hypotheses, often uncovering unexpected cross-domain relationships that would be difficult to detect manually. By integrating structured knowledge from genetics, clinical records, literature, pathways, and curated knowledge bases, knowledge graphs (KGs) enable a more comprehensive and mechanistic understanding of biology and disease [7,8,23–29].

The biomedical knowledge graph (KG) landscape has been described as a “wild west” of inconsistent standards and practices [30,31]. Many KGs are developed with bespoke methods and evaluated using metrics specific to narrow applications, which severely limits interoperability and reuse. This fragmentation leads to redundant reconstruction of many similar graphs, restricts feedback to upstream data providers, and diminishes opportunities for data harmonization and quality improvement. Reuse is further constrained by incomplete documentation, infrequent updates, and insufficient evaluation, factors that collectively undermine confidence and incentivize the creation of new KGs, often as fundable but duplicative deliverables.

Ontologies provide a semantic hierarchical structure and vocabulary for integrating data. Using ontologies in KGs allows complex relationships and links to different concepts, such as genes, traits, or phenotypes, to be represented in a standardized and computable manner [4,32]. For ontology creation, there are defined standards and guidance such as OWL, DOS-DP and ROBOT templates[33–35]. The OBO Foundry provides ontology best practice metrics such as responsiveness, maintenance, format, and others[36]. The Bioregistry is an open source curated database of prefixes and metadata for ontologies, controlled vocabularies, and other semantic types[37]. Bioportal encourages reuse of ontologies through an open source repository[38]. These resources have encouraged rigorous creation and maintenance of ontologies in order to ensure that data can be shared across projects. KGs, built from ontologies and data sources, are now experiencing similar issues as ontologies were before these standards and resources were built. While the ontologies and data sources generally maintain consistent versioning, detailed documentation, and structured formatting, these features are not often retained in the KGs assembled from these parts. It is time that we have the same level of standards for KGs.

Non-biomedical KGs can also provide important lessons in data management and graph creation. Google, Facebook, IBM and others use KGs to organize and deliver information. The size of these KGs necessitates a consistent underlying schema or structure[39]. Without a consistent underlying structure, it would be nearly impossible to create reliable graphs of these magnitudes. Standards in KG development are similarly important in industry and other scholarly domains [40,41], and although use cases vary, the principles are applicable to biological domains as well. Unlike social graphs such as Facebook's, which are largely centered on people and their interactions, biomedical KGs must capture a far more complex landscape. In biomedicine, the same two entities can be linked by different types of relationships depending on the context, and those relationships may change as new evidence is discovered. In some cases, the relationships may even appear contradictory but remain accurate within their specific experimental or biological setting. Additionally, biomedical KGs must also integrate across multiple biological scales (from molecules to cells to organisms to populations) and capture qualifiers such as tissue, species, or regulatory direction. Combined with the need for high accuracy, since an incorrect relation could directly influence scientific or clinical decisions, this semantic diversity and multi-scale integration make their construction and maintenance uniquely challenging.

To address these limitations, a number of community efforts, prominently including contributions from the Monarch Initiative, have developed resources to enhance the accessibility, reusability, and standardization of biomedical KGs. KG-Hub provides community practices and tooling for KG discovery, harmonization, and reuse, in a manner analogous to the role of the OBO Foundry in ontology development [36]. Within this framework, the KG-Registry catalogs metadata describing graph creators, data sources, organizations, and associated code, thereby facilitating identification and reuse of existing KGs[42]. Complementary initiatives include the (Re)usable Data Project, which evaluates and clarifies licensing to promote responsible distribution of biomedical data—an important consideration given that KGs typically integrate sources with heterogeneous reuse conditions[43]. The Biolink Model is a standardized framework for representing biological entities, their attributes, and relationships[44]. Adoption of Biolink has been critical to ensuring interoperability, and associated standards such as KGX exchange format leverage Biolink to enable subgraph extraction and integration across resources. Broader community efforts, including the NCATS Biomedical Data Translator Consortium[45], and Translator KGs such as RTX-KG2, ROBOKOP, Clinical KG, and SPOKE[1,46–48], demonstrate how strong, schema-driven approaches based on shared formats and registries provide the foundation for disseminating standardized and reusable KGs.

Versioning is critical for the reliability and reuse of KGs. Lessons from ontologies illustrate this need: a longitudinal study of the Gene Ontology (GO) showed that annotation changes over a decade produced divergent interpretations of biological experiments, with enrichment results differing substantially between earlier and later GO releases[49]. Such discrepancies arose from biases including annotation and literature bias, underscoring the dynamic nature of biological data and the necessity of regular updates. Outdated KGs risk propagating obsolete or misleading results, particularly when based on stale source data. To support reproducibility, it is desirable for past KG versions to remain accessible for comparison with newer iterations. However, practical barriers—most notably the large size of KGs and the resource demands of hosting and bandwidth—make comprehensive version archiving difficult.

Guidance can be drawn from standards used in other data domains. The FAIR, TRUST, and O3 principles provide frameworks for sustainable, interoperable data resources[50–52]. FAIR is Findability, Accessibility, Interoperability, and Reusability, and its recommendations include use of controlled vocabularies for metadata, schematic interoperability across sources, and modular design to enhance flexibility. Vogt et al. (2024) argue that modularity-oriented approaches to KG construction improve compatibility, findability, and reproducibility[50]. TRUST principles extend FAIR by emphasizing long-term stewardship of digital repositories, while O3 highlights sustainability through open data, code, and infrastructure. Yet, none of these frameworks fully address the unique requirements of KG assembly and reuse. For this reason, we defined criteria tailored to KGs and applied them to a set of prominent, publicly available graphs.

Several biomedical KG reviews have focused on KG use cases and methods; but not their creation and reusability [6,53–55]. We were inspired by *Bonner et. al [54]*, where they examined schema, relations and data mappings, and licensure for six drug discovery KGs, but not KG evaluation and examples on their intended purpose. They concluded that three main things missing from these KGs are sufficient documentation, regular updates, and consistent dataset versioning. In response, we present a systematic approach and set of KG metrics designed to improve interoperability, rigor, reuse, and reproducibility. By addressing these foundational issues, we aim to foster more collaborative and modular development across the community and to enable researchers to focus on applying KGs to drive new discoveries and achieve real-world impact.

# Methods

Here, we review a subset of 16 biomedical KGs[1,7–10,23,46–48,56–62] with a focus on reusability, data standards and openness. All metrics are evaluated through the lens of "*is it publicly accessible/true?"* Features only accessible through subscription or membership in a specific group are not counted. The features were assessed as a strictly *"yes"* or *"no"* system; partial compliance counted as a full *"yes."* Only features that were completely missing or could not be found were counted as a *"no."* Due to the variation in how KGs define, represent and approach concepts we kept our criteria intentionally broad to avoid penalizing approaches following different standards. Our goal was to give credit for efforts towards implementing or adhering to standards.

KGs were evaluated manually by domain-expert curators addressing specific subcriteria for each Principle (see below). For each KG, one of a team of six reviewers retrieved the most recently available version of the KG's website as well as the contents of any associated software repositories (e.g., from GitHub) and scholarly publications. The review was then secondarily reviewed by another team member. A critical review of KG code and files is outside the scope of this work and was not performed (e.g., we did not determine how much time was required to build any KG). As with the OBO Dashboard and the ReusableData initiatives[36,43] 24,33, it is our hope that this review process will encourage KG stewards to improve their documentation and promote reuse and interoperability.

# Principles

We developed specific evaluation criteria to assess the accessibility, reusability, and standardization of publicly-accessible KGs. The primary goal of these principles is to determine if one can easily access and gain specific information about a given KG so as to determine its utility for reuse. Our criteria are arranged according to six principles:

A. Access Level and Type
B. Provenance of Nodes and Edges
C. Documented Standards, Schema, and Construction
D. Update Frequency and Versioning
E. Evaluation by Metrics and Fitness for Purpose
F. Licensing

## A. Access Level and Type

We evaluated each KG's ability to be accessed easily and without barriers by members of the greater scientific community. We considered direct access to the KG and its code as well as adjacent interfaces (e.g., a REST API) and derived products (e.g., embeddings, paths). The subcriteria are:

A.1 Additional KG product accessible such as embeddings or paths
A.2 API, Neo4j or other online hosted graph database accessible
A.3 Multiple ways to download or access
A.4 Source code for making the KG is publicly accessible
A.5 KG is downloadable in its entirety

## B. Provenance of Nodes and Edges

KG sources can vary noticeably in structure and information content. Having access to original data sources can help users understand the importance of different edges and nodes, increasing biological plausibility and classification for drug repurposing tasks[63]. Knowledge of data sources can also inform choices about a KG's reusability: some data sources may be more or less stringent than a user may prefer, so documentation of the process from source to KG can help users to evaluate the graph's suitability for new purposes. The subcriteria are:

B.1 List of sources provided
B.2 Sources have versions/dates/file names used
B.3 Import dependencies declared
B.4 Nodes and edges have source information
B.5 Duplicate edge management taken into consideration
B.6 Provenance of edge creation
B.7 Schema for how edge types are created; where are they coming from

## C. Documented Standards, Schema, and Construction

KGs gain much of their value from being reusable across different applications and research contexts. For this to happen, the KG's identifiers need to be resolvable in the greater biomedical domain. In addition, how the graph is created may change the way a user employs it. Here, we assess how the documentation of the KG best supports this. The subcriteria are:

- C.1 Entities and nodes are human readable and usable in biological applications outside of KG
- C.2 Identifiers support entity resolution (external identity links such as Ensembl IDs[64])
- C.3 Documentation on construction is provided
- C.4 Documented data transforms (for example, removed general terms from ontologies)
- C.5 Uses a documented schema for construction

## D. Update Frequency and Versioning

As KGs are created from current "known" knowledge, they risk missing new information if not updated regularly; timely, well-documented updates are necessary. This is analogous to software versioning in that knowledge of the version used in a given process encourages reproducibility and more informative evaluation of the resource's reuse in new applications. The subcriteria are:

- D.1 Clearly identified stable versions
- D.2 Public tracker for requests
- D.3 Contact indicated for the KG; either organization or person
- D.4 Updated more than once per year
- D.5 Prior versions accessible with clearly documented changes

## E. Evaluation by Metrics and Fitness for Purpose

Many KGs claim they will help scientists in the broader biological community in a spectrum of use cases. Formal evaluations, detailed examples, and quantifiable measures of confidence are necessary to establish the community's trust in the KG's value and applicability to its intended use cases. The subcriteria are:

- E.1 Case study/example of use provided
- E.2 KG utility evaluated against other methods (can include other KGs)
- E.3 Clearly defined scope
- E.4 Multiple types of evaluation methods
- E.5 One or more ways of measuring accuracy/confidence

## F. Licensing

Knowledge Graph data licensing is important because it clearly defines what users are legally allowed to do with the data, such as reuse, redistribution, or modification. Without proper licensing, even openly available data may not be legally reusable, creating barriers to collaboration and scientific innovation. Clear licenses enable interoperability, sharing, and the creation of tools and resources that advance research and improve human health.

- F.1 KG licensing is clearly stated and unambiguous

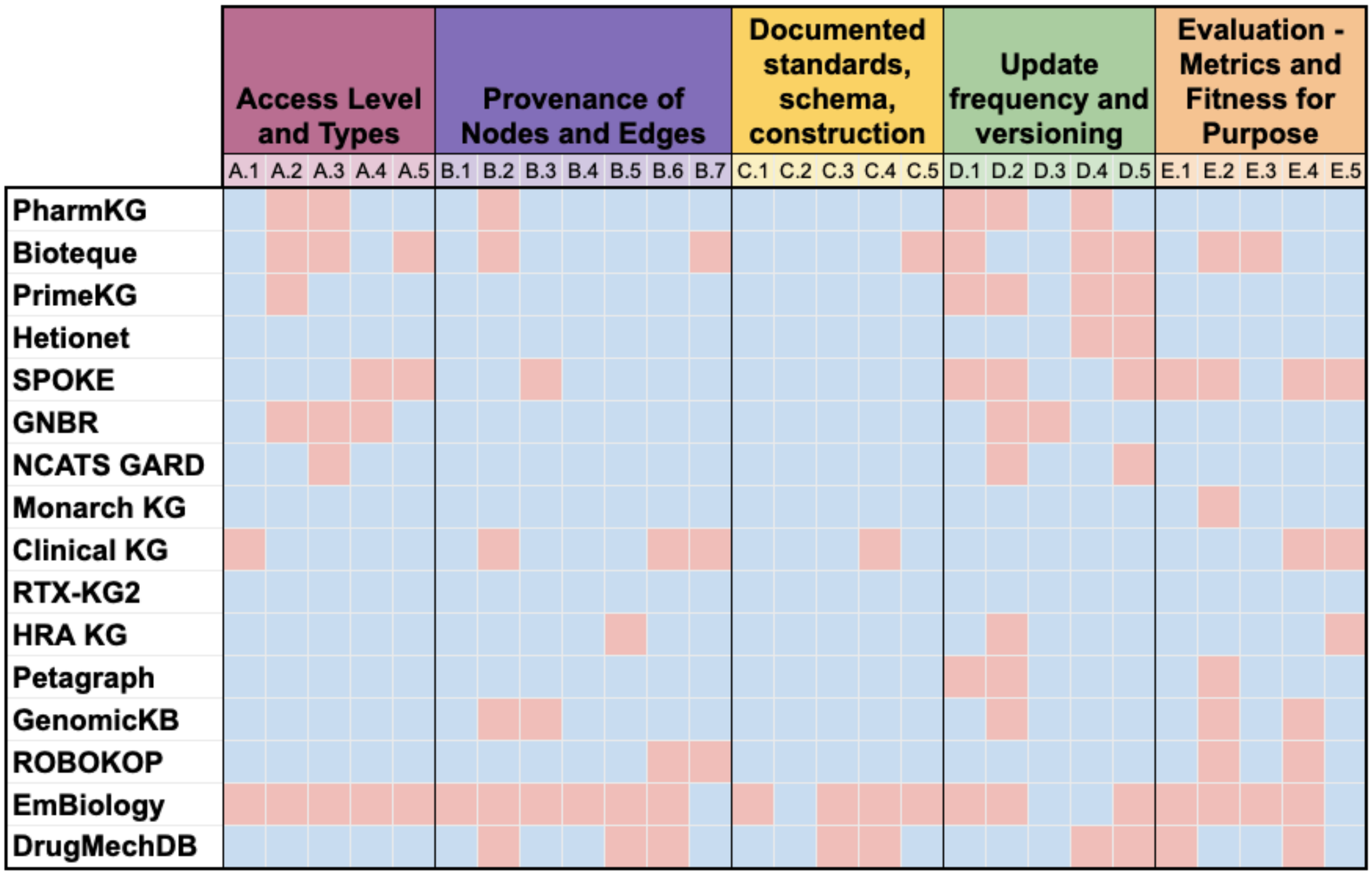

**Figure 1.** The left side lists the KGs reviewed.The top indicates the different criteria by category with each color referring to a different category. The criteria are numbered A.1, B.1,.... according to the numbering above. Blue squares indicate the KG met this principle, red indicates it did not. Licensing is not shown as it is not a binary category. Find a detailed breakdown of why each KG was assigned blue/red in the **Supplemental Table Detailed Principle Analysis** tab.

# Results

## KG Evaluation

The sixteen chosen KGs are shown below in **Figure 1** along with how they did on the criteria from above. We chose biomedical KGs of varying purposes, sizes, and creators, including those that have been widely cited such as Hetionet and Bioteque and those already in our KG-Registry. The reviewers for the Monarch Initiative's KG, were experts outside of the Monarch Initiative in order to reduce bias.

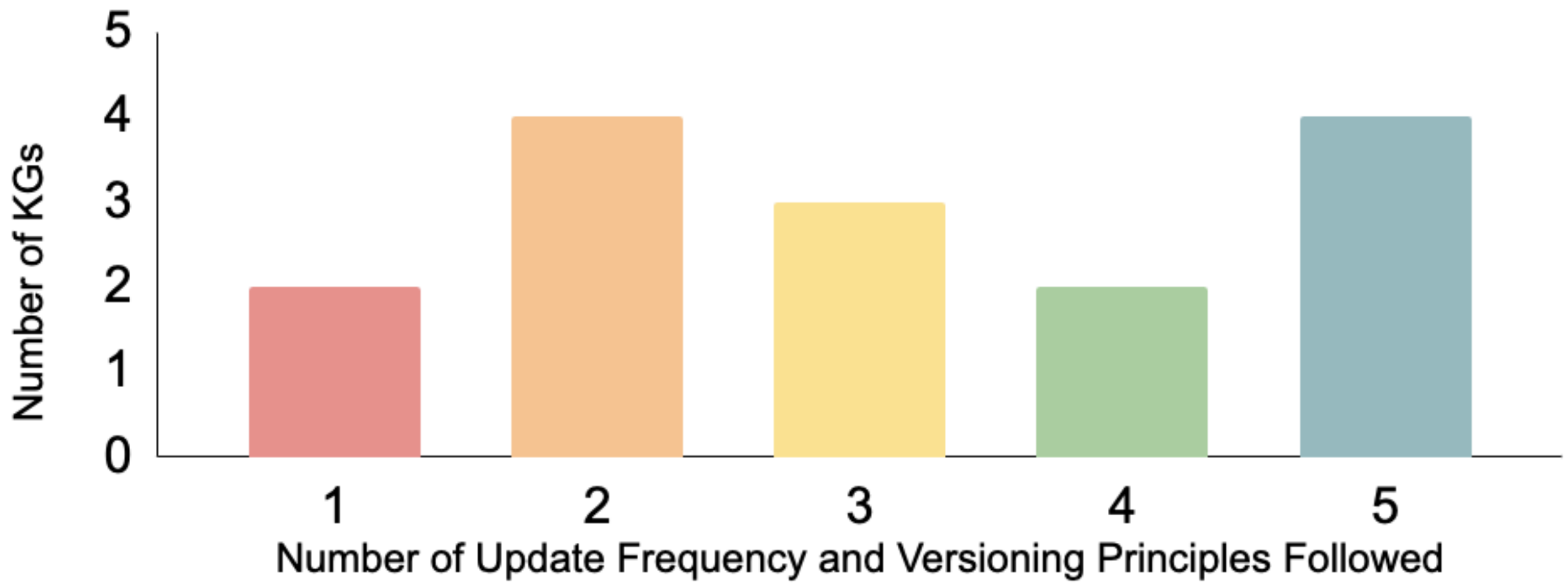

**Figure 2.** The x-axis shows the number of update frequency and versioning principles that a KG followed out of a five. The y-axis shows the number of KGs that we reviewed for each criteria in category D.

Of note, KGs such as EmBiology and SPOKE are largely accessible only through a paywall, which has resulted in lower scores. RTX-KG2 was the only knowledge graph that scored yes on every criteria.

**Figure 2** shows how many KGs follow principles of update frequency and versioning. Only Monarch KG, Clinical KG, and RTX-KG2 scored yes on all update frequency and version principles. All KGs except for GNBR have a contact person or organization indicated. The principles which were followed the least are having a public tracker for requests (7/16) followed by clearly identified stable versions and having prior versions accessible with clearly documented changes (9/16). This creates a lack of transparency in how KGs are maintained and improved, and contributes to problems with reproducible workflows.

## KG Nodes Types

From the 16 KGs we reviewed the number of node types included ranged from many (52) to very few (3) with many KGs falling somewhere in between (**Figure 3)**. Common node types were chemicals/drugs, genes and diseases. Some KGs had more custom node types such as HRA KG which used digital objects (DOs) instead of standard nodes. This design enables HRA KG to incorporate experimental data, research settings, and even 3D spatial references, giving it a distinct focus compared with KGs that limit themselves to curated biomedical entities such as genes, proteins, and diseases, without experimental data. In the case of EmBiology, publicly available information is limited: based on their website we infer that their KG includes genes, proteins, and diseases, but the details remain inaccessible because of a paywall. To better compare what kinds of nodes are used across KGs, we leveraged ChatGPT to map each node label supplied by the different KGs to the Biolink Model's node types. The full list is provided in the **Supplemental Table Node Types** tab**. Figure 3B** shows how the different KGs cluster, according to seaborn clustermap, when node types have been mapped to BioLink. For example, the node type *"Tissues",* present in GenomicKB, Bioteque and NCATS GARD was mapped to *"anatomical entity"*, these KGs did not have node type *"anatomical entity"* in their original list.

This mapping allows us to compare KG content as RTX-KG2, ROBOKOP and Monarch all had *“anatomical entity”.*

## KG Sources

Just as the number of node types varied, the number and diversity of sources also differed widely across KGs, encompassing both primary sources (original databases or ontologies) and aggregated sources (resources that themselves integrate multiple databases). As most KGs did not differentiate between primary and aggregated sources they are counted equally. Of the KGs that we evaluated (**Figure 4)**, Monarch ingested the most number of sources at 87, with the next highest being RTX-KG2, ROBOKOP and SPOKE with 63, 50 and 47, respectively. PharmKG and GNBR had the fewest sources with seven and one. Due to EmBiology’s paywall it is unclear exactly how many sources they ingest, although we can deduce there are at least six. GNBR is a literature based knowledge graph, although constructed from many sources (papers) we classify it as having one source “other”. Some KGs included experimental data or text mined data not found in a database, which we classified under the source type “other.”. KGs that included “other” source data include ROBOKOP, Bioteque, DrugMechDB, EmBiology, and GNBR.

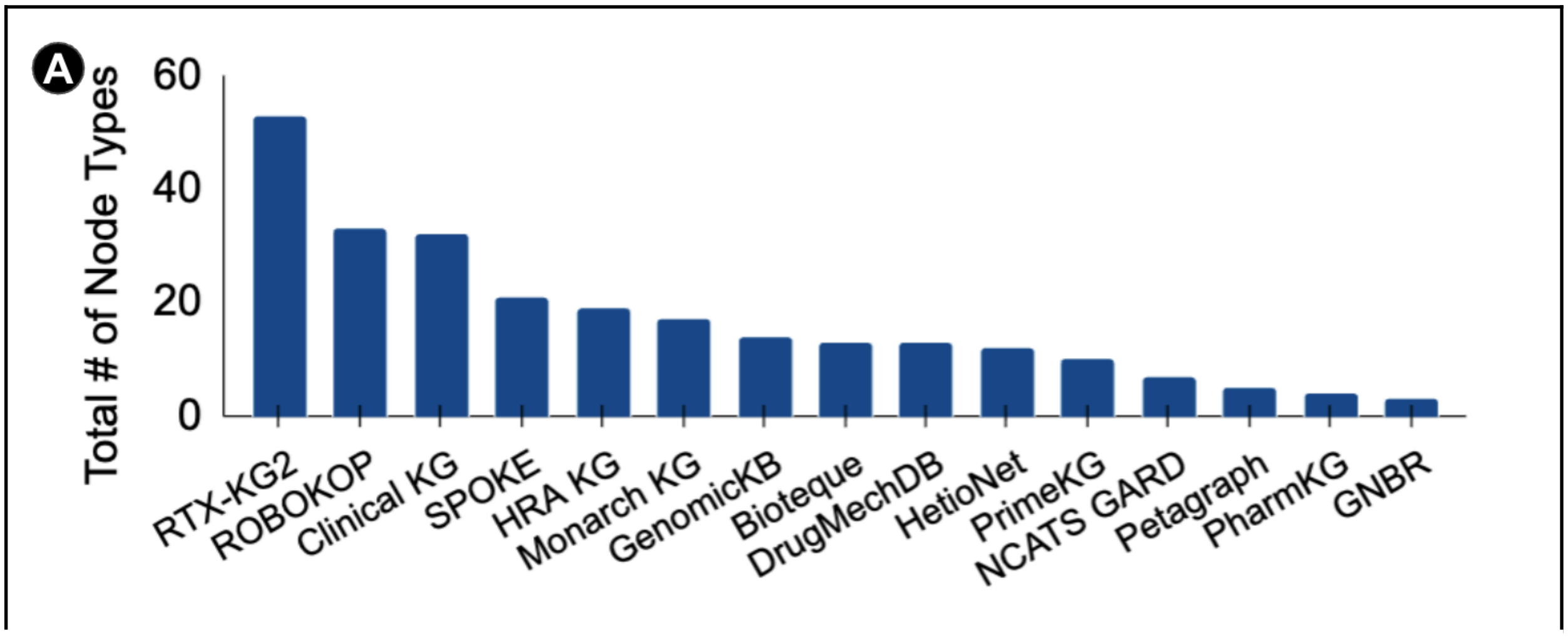

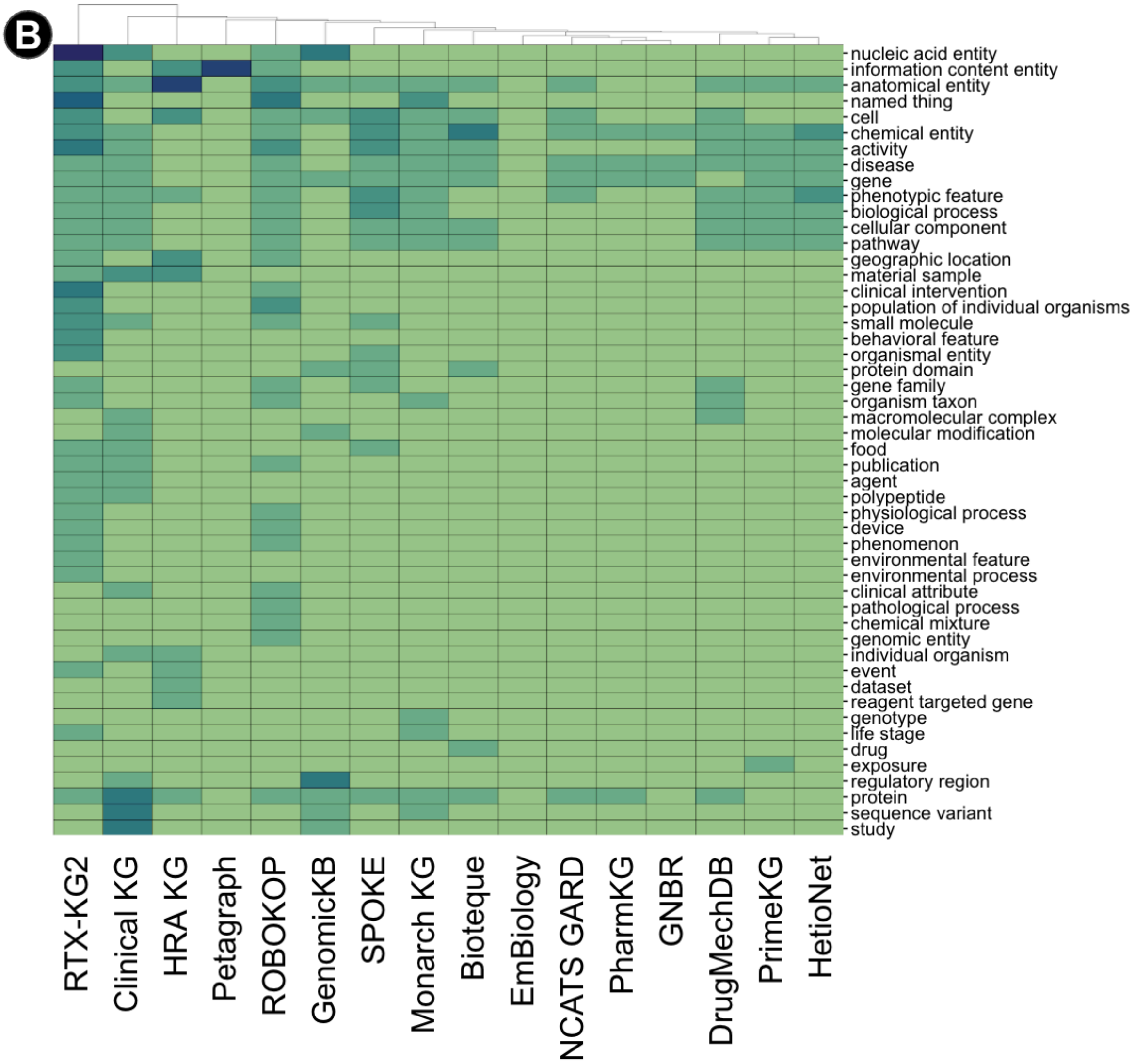

**Figure 3. (A)** X-axis shows the KG, y-axis shows the number of node types reported by each respective KG. EmBiology is not included in this figure as their node types are not publicly available. See **Supplemental Table Node Types** tab for more info. **(B)** Heatmap showing the KGs and their node types after BioLink mapping. The columns represent the KGs, the rows represent the BioLink node types. Darker blue indicates KG has more of the specific node type while green indicates the KG does not have the node type. Dendrogram across the top shows which KGs are more similar based on binomial clustering.

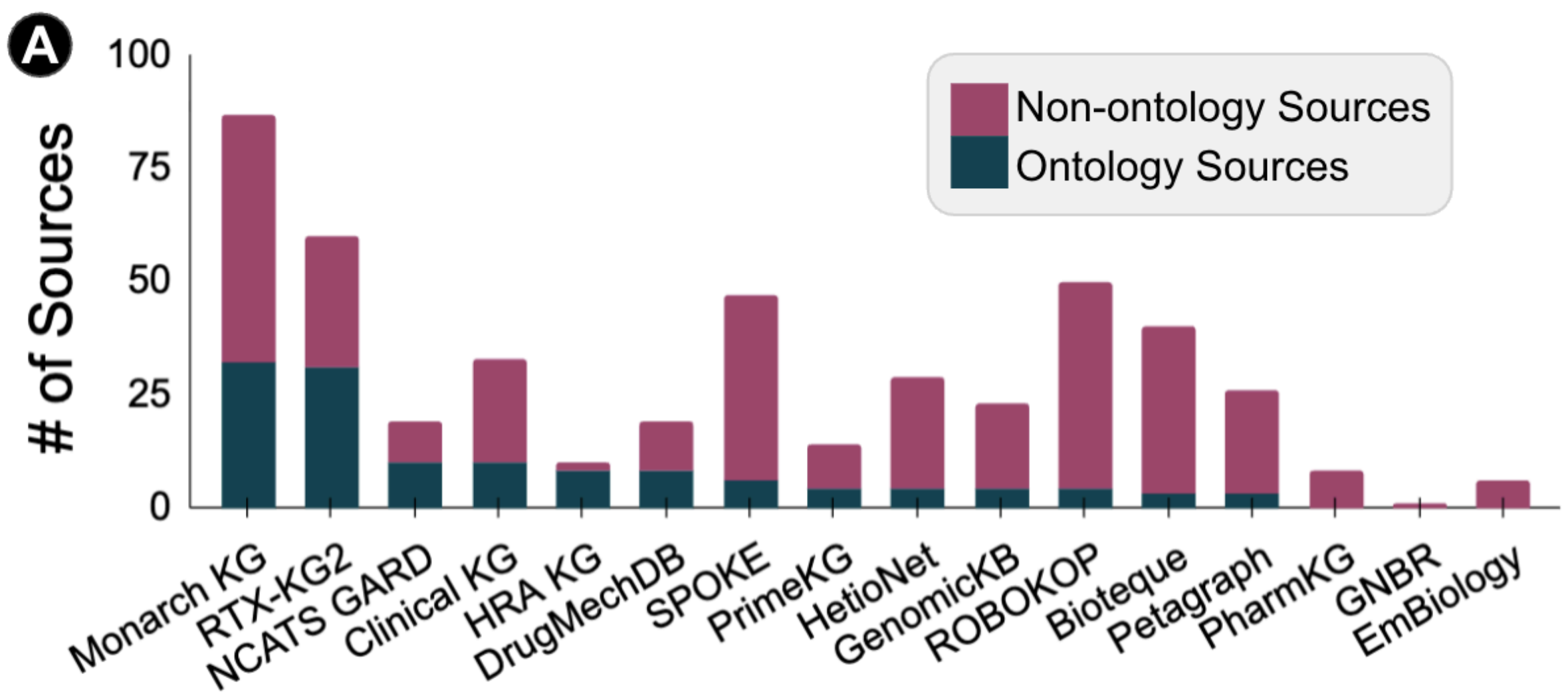

**B** Top 10 most used Ontology Sources

| | GO | HPO | Cell Ontology | Uberon | ChEBI | OMIM | DO | MeSH | MONDO | BTO |
|---|---|---|---|---|---|---|---|---|---|---|
| RTX-KG2 | ✓ | ✓ | ✓ | ✓ | ✓ | ✓ | ✓ | ✓ | ✓ | |
| SPOKE | ✓ | | ✓ | ✓ | | ✓ | ✓ | ✓ | | |
| ROBOKOP | ✓ | ✓ | | ✓ | ✓ | ✓ | | | ✓ | |
| DrugMechDB | ✓ | ✓ | ✓ | ✓ | ✓ | | | ✓ | | |
| HetioNet | ✓ | | | ✓ | | | ✓ | ✓ | | |
| Monarch KG | ✓ | ✓ | ✓ | ✓ | ✓ | ✓ | | | ✓ | |
| Bioteque | ✓ | | | | ✓ | | | | | ✓ |
| Clinical KG | ✓ | ✓ | | | | | ✓ | | | ✓ |
| PrimeKG | ✓ | ✓ | | | | | ✓ | | ✓ | |
| GenomicKB | ✓ | | ✓ | ✓ | | | | | | ✓ |
| NCATS GARD | | ✓ | ✓ | | ✓ | ✓ | | ✓ | ✓ | |
| PharmKG | | | | | | ✓ | | | | |
| HRA KG | | | ✓ | ✓ | | | | | | |
| Petagraph | | ✓ | | | | | | | | |
| EmBiology | | | | | | | | | | |
| GNBR | | | | | | | | | | |

**C** Top 10 most used Non-Ontology Sources

| | Drugbank | Reactome | DisGenNET | Drug Central | SIDER | UniProt | STRING | BGee | GWAS Catalog | HGNC |
|---|---|---|---|---|---|---|---|---|---|---|
| RTX-KG2 | ✓ | ✓ | ✓ | ✓ | | ✓ | ✓ | | | ✓ |
| SPOKE | ✓ | ✓ | ✓ | ✓ | ✓ | ✓ | | ✓ | ✓ | |
| ROBOKOP | ✓ | ✓ | ✓ | ✓ | ✓ | ✓ | ✓ | ✓ | ✓ | ✓ |
| DrugMechDB | ✓ | ✓ | | ✓ | | ✓ | | | | |
| HetioNet | ✓ | ✓ | ✓ | ✓ | ✓ | | | ✓ | ✓ | |
| Monarch KG | | ✓ | | | | ✓ | ✓ | ✓ | | ✓ |
| Bioteque | ✓ | ✓ | ✓ | ✓ | ✓ | | ✓ | | | |
| Clinical KG | ✓ | | ✓ | | ✓ | ✓ | ✓ | | ✓ | |
| PrimeKG | | ✓ | ✓ | ✓ | ✓ | | | ✓ | | |
| GenomicKB | | | | | | | | | ✓ | |
| NCATS GARD | | | | | | | | | | |
| PharmKG | ✓ | | | | ✓ | | | | | |
| HRA KG | | | | | | | | | | ✓ |
| Petagraph | | | | | | ✓ | ✓ | | | ✓ |
| EmBiology | ✓ | | | | | | | | | |
| GNBR | | | | | | | | | | |

**Figure 4. (A)** x-axis shows the KG, y-axis shows the corresponding number of sources the KG ingests. Sources have been divided into two categories: ontologies (in purple) and non-ontology data sources (in blue). **(B)** The right side of the table shows the different KGs, the top indicates the top 10 most ingested ontology sources. A blue check indicated the KG ingested the source. **(C)** The right side of the table shows the different KGs, the top indicates the top 10 most ingested non ontology sources. A blue check indicated the KG ingested the source. More information about the KG Sources can be found in the **Supplemental Table Sources** tab**.**

# Discussion

When a researcher is deciding to use a knowledge graph for an analysis or experiment, they face a decision: to create a custom graph tailored to their purpose or use an existing resource and try to adapt it to fit their use case. There are advantages and disadvantages to each approach; however, given the rapidly growing number of biomedical KGs and lack of apparent reuse, it appears that there is a tendency to create one’s own. There are several reasons for this, despite the high overlap in content and sources across existing graphs. First, it is often considered easier to make something from scratch and fully understand how it works and ensure compatibility with existing or planned workflows. This need for understanding is why documentation on KG construction is so important - and based on our evaluation, often not sufficient - to convince users to trust the creators enough to reuse. A commonly encountered challenge is that the necessary information is not consolidated or easy to find. When assessing a KG and its principles, the reviewers often had to look in multiple places such as publications, websites, and github repositories to find the relevant information. Even if the KG scored ‘yes’ on a principle does not mean it was easily scored. Second, creating novel KGs is more compatible with the way research is funded and rewarded. It may be more impressive to say “we made a thing” than it is to say “we adapted this existing data and framework”, unless there is prior evidence that reuse has a measurable benefit, research tends to follow funding priorities that incentivize creation rather than reuse or validation. As such, there has been a steady stream of similar KGs being built. While this expansion might seem productive, it often comes at the expense of exploring novel methods that could make better use of existing KGs or evaluation of applying existing KGs to novel areas of exploration. The result is a cycle of effort duplication rather than innovation. If the field eventually turns away from KGs due to limited meaningful results, the real loss will be the missed opportunity to advance methods and applications that could have demonstrated their value. By continually rebuilding what we already have, we risk exhausting interest before KGs reach their full potential.

As a community, we are not generating the evidence needed that reuse actually helps advance the field because of the lack of trust in any given KG. This paper aims to help build a set of criteria to evaluate the trustworthiness of KGs in order to help researchers better build upon each other's work. Each KG was assessed against our criteria with a simple “yes/no” scoring system, where the number of “yes” answers provides an overall score of transparency and reusability. Well documented code and explicit descriptions of the data structure encourages methodological innovation by computational biologists. Additionally, being able to contact the authors to ask for clarity or suggestions offers another layer of trust. A good way to build trust would be for KGs to have a machine readable overview file dictating things about the KG such as date updated, authors, size, sources, etc. In this study, we do not set a strict cutoff for what constitutes a “trusted” KG (e.g., requiring a minimum number of “yes” scores), but we propose that higher scores across these criteria reflect greater trustworthiness. We also report scores across the criteria in **Figure 1** to enable head-to-head comparison.

Even if users do trust existing KGs, the next question is which one do they pick? Not all KGs are created equal for all biomedical tasks. In order to confidently choose a KG or set of them, researchers need to know the different data types in the graph and the data sources used to build it. As shown in **Figure 3 and 4,** KGs have a varying number of node types and sources ingested. When an outside user looks to repurpose an existing KG, it needs to be clear which information is already represented and where it came from. Data transparency enables a user to take an existing KG and add or remove their own data sources in a compatible way, making it easier to repurpose. **Figure 4** shows that many KGs ingest similar sources. However, this does not guarantee that the content is of equal quality once it is aggregated into a KG. An equally important aspect of KG trustworthiness is the accuracy of resource ingestion. Even when two

KGs draw from the same source, they may ingest and represent the dataset in different ways. For instance, while both ROBOKOP and Monarch KG ingest some of the same sources, they often pull from different tables within those sources, leading to differences in coverage and emphasis. In another instance, PharmKG focuses on diseases above the fifth hierarchical level in MeSH while Monarch KG makes use of all diseases in MONDO. While cutting more general terms could be useful in specific tasks it would lose information content inherent to the source ontologies. Knowing these KG design choices, influences the choice of both KG and methods to use for a specific use case.

Embedding KGs may be a useful strategy for data that cannot be publicly available, such as sensitive clinical data. However, without excellent documentation, frequent updates, and robust evaluation and validation, embeddings as the sole shareable artifact have little practical value.. Embeddings are constrained to the version of information and embedding strategy used at the time of creation. There is no way for an outside user to add new information or generate embeddings using alternative techniques. Additionally, embeddings also lose the data providence and explainability that are inherent to well constructed KGs. We urge the community to further define best practices and approaches to sharing and validating embeddings, much in the same way that is currently done for bringing models to data[65,66].

When evaluations of biomedical KGs are provided, they vary greatly in scope, application and rigour. More general evaluations of a KG, such as agreement with other sources, can increase trust in the information contained, but do not provide insights into how the KG may be useful for specific analysis. Conversely, some evaluations only consider a specific use case, for example, insights into a specific disease (e.g., Alzheimer's, epilepsy) or drug. This level of specificity inherently limits what the KG is likely to be used for in the future. A robust analysis that includes both general domain-agnostic evaluations and nuanced use-cases comparing a novel KG to existing methods or knowledge sources will generate more trust and opportunities for reuse.

When assessing performance of KGs on machine learning applications like drug repurposing, fairness in evaluation is critical. Using a validation set drawn from one of the KGs inevitably skews results towards that KG, since the topological structure of each KG differs. This bias is particularly pronounced in machine learning tasks such as link prediction, where the KG that supplied the validation set will closely resemble the topology of the known connected pairs. The other KG, by contrast, may appear to underperform simply because its structure reflects different sources and ingestion protocols. To mitigate this bias and enable more meaningful comparisons, external validation sets curated independently of either KG should be used.

Many of the KGs have overlaps in sources and node types, as expected. However, there is no guarantee that the node type in *KG (a)* is the same as node type in *KG(b)* even if they have the same name. Conversely, node type ‘a’ in *KG(a)* may have a different name than node type ‘b’ in *KG(b)* but be essentially the same data. The same source can also be ingested differently making two seemingly similar graphs more different than one would expect. Two KGs can ingest the same data but from two different places such as a primary data source (the original ontology) vs an aggregated source that has many ontologies. For example, one can ingest the primary source such as the Basic Formal Ontology (BFO) as opposed to OBO Foundry which ingests many ontologies including BFO. Primary vs aggregated source reporting could make two KGs look different on the surface when in fact they are quite similar. As KGs follow different standards and data models, there is not a good way to compare the sources and nodes in different KGs in a robust non biased way. Currently it is nontrivial to compare different knowledge graphs however, if all KGs followed similar construction methods and were in the same format such as KGX with the same metadata stored in machine readable files, comparison would not only be easier, but standard.

# Conclusion

In conclusion, for KGs to reach their full potential in the biomedical landscape we must shift from the fragmented landscape of reinvention and come together as a community towards collective stewardship. Strengthening existing KGs through rigorous documentation, transparent standards, and shared evaluation frameworks will make reuse not just possible, but preferable to building yet another graph with overlapping content. By embracing common principles of construction and machine-readable metadata, researchers will be empowered to compare, adapt, and extend KGs with confidence. The result will not only be greater trust in individual KGs, but also a fertile ecosystem where innovation flourishes because energy is spent advancing methods and applications, rather than duplicating what already exists. Only by uniting around shared standards can we ensure that KGs mature into reliable, reusable, and indispensable infrastructure for the biomedical sciences.

# Funding

This research is generously supported by the

- National Institutes of Health (NIH) Office of the Director Grant #5R24OD011883
- NIH National Human Genome Research Institute Phenomics First Resource (NHGRI) #7RM1HG010860
- NIH #5T15LM009451-19
- NIH R35 GM128765
- Simons Foundation 1017799

# Detailed Principle Analysis

| | A. Access Level and Types | | | | | B. Provenance of Nodes and Edges | | | | | | | C. Documented standards, schema, construction | | | | | D. Update frequency and versioning | | | | | E. Evaluation - Metrics and Fitness for Purpose | | | | | F. License Information |
|---|---|---|---|---|---|---|---|---|---|---|---|---|---|---|---|---|---|---|---|---|---|---|---|---|---|---|---|---|
| | A.1 | A.2 | A.3 | A.4 | A.5 | B.1 | B.2 | B.3 | B.4 | B.5 | B.6 | B.7 | C.1 | C.2 | C.3 | C.4 | C.5 | D.1 | D.2 | D.3 | D.4 | D.5 | E.1 | E.2 | E.3 | E.4 | E.5 | |
| | Can Access something that isnt original KG (embedding, paths) | Can access by API or Neo4j or other online hosted graph database | Multiple ways to download/ access | Can access source code for making KG | Can download KG | List of sources provided (record sources if yes) | Sources have versions/dates/file names used | Import dependencies declared | Nodes and Edges have source information | Duplicate edge management taken into consideration | Source specifics for what triples are created | Schema for how edge types are created; where they are coming from | Data is human readable and usable in biological applications outside of KGG | Identifiers are resolvable for entity resolution (external identity links such as Ensembl IDs) | Documentation on construction is provided (did they make an effort?) | Documented data transforms (ex: removed general terms from ontologies | Uses a documented schema for construction | Clearly Identified Stable Versions | Public tracker for requests | Is there a contact indicated for the KG either org or person | Updated more than once per year | Prior versions accessible with clearly documented changes | Do they have a case study/example of use | KG utility evaluated against other methods (can include other KGs) | Clearly defined scope | Multiple different evaluation methods | One or more ways of measuring accuracy/ confidence | |
| **PharmKG** | Yes - PharmKG provides embeddings, learned low-dimensional representations for entities and relations via KGE models. It also provides top predicted paths for mechanistic interpretation (Fig 6) and t-SNE visualizations showing semantic structure (Fig 5). | N | N | Yes - The full source code for KG construction, processing, and the Heterogeneous Graph Attention Network (HRGAT) embedding model is provided on the GitHub repository https://github.com/MindRank-Biotech/PharmKG/tree/master | Yes - The benchmark KG is downloadable. They also provide both the clean version (final benchmark) and the raw version (with more entities and relations); the KG dataset is guarded by Google Drive permissions to access it | Yes — An integration of 6 sources (OMIM, DrugBank, PharmGKB, TTD, SIDER, HumanNet); plus entity features from MeSH, PubChem, BioBERT, BioGPS, and Connectivity Map | No - Do mention integrating "recent versions" of each source and give details of ID mapping and unification (e.g., Entrez Gene ID, MeSH hierarchy) but no explicit version numbers or file names for each data dump are shown in the text | Yes - The code is built on Pykeen and KG-reeval, with details for training and hyperparameters. Also mentions use of RDKit for chemical fingerprints and BioBERT for embeddings | Partially - Confirms that nodes were unified with standard IDs (Entrez, MeSH, PubChem) and that duplicate synonyms were resolved, but it doesn't state that each edge carries explicit source provenance tags — only that relations were merged and thematically assigned using Global Network of Biomedical Relationships (GNBR) | Yes - Explicitly describes merging overlapping triplets, disambiguation with synonym tables, and merging semantically similar relations using clustering | Yes - Explains how 29 relation types were derived, how Global Network of Biomedical Relationships (GNBR) semantic themes were mapped, and which sources contribute to which entity types and interactions | Yes - Documents edge semantics: "interaction," "disease-gene," "chemical-gene," "disease-chemical," and subtypes — all mapped from Global Network of Biomedical Relationships (GNBR) semantic themes and curated bases | Yes - All entities explicitly use standard biomedical identifiers and the structure is designed to support drug repurposing, target discovery, adverse reaction prediction, etc., in real-world biomedical tasks | Yes - Genes use Entrez Gene IDs; diseases use MeSH IDs; chemicals use PubChem IDs | Yes - Explains the construction in detail, including entity filtering, merging, disambiguation, feature extraction, and the final schema design | Yes - Removed trivial entities, merged low-level symptoms into MeSH parent diseases, clustered and merged redundant relation types, applied PCA for feature reduction, and explain all steps clearly | Yes - 29 defined relation types with source mapping and entity categories; and semantics from Global Network of Biomedical Relationships (GNBR) themes and curated bases | No - There is no formal semantic versioning scheme mentioned for the dataset or codebase | No - There is no mention of a public tracker for feature requests or issues | Yes - Two corresponding authors are provided with full institutional contacts (Prof. Yuedong Yang and Dr. Zhangming Niu) in the published paper (https://academic.oup.com/bib/article/22/4/bbaa344/6042240) | No - No evidence yet. They state they plan future expansions (e.g., transcriptomics, clinical data, auto-extraction) but no recurring updates are published so far | Partially - They do provide the raw version alongside the final benchmark on GitHub but do not maintain a changelog across multiple yearly versions | Yes - Provide detailed case studies for Alzheimer's disease and Parkinson's disease for drug repurposing and target discovery, with literature validation, top scored predictions, and visualized paths | Yes - Benchmarked Heterogeneous Graph Attention Network (HRGAT) and 9 other KGE baselines on Hetionet and PharmKG side by side and analyzed results | Yes - The paper is a dedicated benchmark for evaluating KGE models in biomedical relation prediction, with explicit focus on drug repurposing, target identification, and multi-relation prediction tasks | Yes - Link prediction (MRR, Hits@k), downstream tasks (AUROC, AUPRC), t-tests for significance | Yes - Multiple: MRR, Hits@1/3/10/100, AUROC, AUPRC, p-values for statistical tests, plus visualization and interpretability checks with t-SNE plots and path analyses | No explicit license (restricted use). The paper notes: "© The Author(s) 2020. Published by Oxford University Press. All rights reserved." But does not reveal license information (nor does the GitHub repo) |
| **Bioteque** | Y (Bioteque provides node-type-specific embeddings for 11 of the node types in its graph, along with downloadable sets of those individual sets of nodes mapped to their representations in the embeddings.) | N | N | Y (Source code for assembling the KG is provided in a GitHub repository, https://github.com/sbnb-irb/bioteque. This includes source-specific scripts for retrieving data components.) | N | Y (see https://bioteque.irbbarcelona.org/sources) | N | Y (The list of data sources is documented and includes specific mentions of source files in its data retrieval scripts, e.g., the source for COSMIC is defined here: https://github.com/sbnb-irb/bioteque/blob/master/datasets/cosmic_mutsig/get_data.sh) | Y (The sources of nodes and edges are provided in the documentation and the node files.) | Y (The 2022 Bioteque paper makes two notes about duplicate resolution: "...we first mapped the samples and genes to our reference vocabulary and collapsed the duplicates by their mean value", and "We mapped the cell lines and genes to our reference vocabularies and took the mean value whenever duplicates occurred".) | Y (The list of data sources is documented and makes specific mention of which associations are derived from which sources; see https://bioteque.irbbarcelona.org/sources) | N | Y (the provided node files are provided as TSVs, though assembly of a full KG would require running the graph assembly code. It appears that the assembly code also produces nodes and edges in a TSV format.) | Y (node identifiers are from clearly defined sources and expressed as CURIEs) | Y (Documentation regarding the assembly code is provided; see https://github.com/sbnb-irb/bioteque) | Y (each source has its own transform code and documentation, provided on the GitHub repo, https://github.com/sbnb-irb/bioteque) | N | N | Y (the GitHub repository at https://github.com/sbnb-irb/bioteque is public and permits issue creation). | Y (it's never explicitly stated as a contact, but the responsible organization, the Structural Bioinformatics and Network Biology Group at the Institute for Research in Biomedicine Barcelona, is identified along with a link to their home page) | N | N | Y (Examples of use are described in the 2022 Nat Comm paper; an example of generating predictions for drug repurposing is provided.) | N | N | Y (Multiple evaluation methods are provided in the 2022 Nat Comm paper, primarily for embedding evaluation.) | Y (Multiple validation methods are provided in the 2022 Nat Comm paper, including two distinct analyses involving gene expression data and protein-protein interactions, respectively.) | CC BY 4.0 |
| **PrimeKG** | Yes - ClinicalBERT-based embeddings were used to group disease nodes, providing an embedding-derived version of the graph | N | Yes - Available via Harvard Dataverse with raw KG (kg raw.csv) and largest connected component (kg giant.csv) https://dataverse.harvard.edu/dataset.xhtml?persistentId=doi:10.7910/DVN/IXA7BM | Yes - Full source code is available on GitHub https://github.com/mims-harvard/PrimeKG | Yes - Harvard Dataverse Repo hosts the downloadable KG and intermediate files https://dataverse.harvard.edu/dataset.xhtml?persistentId=doi:10.7910/DVN/IXA7BM | Yes - 20 primary data sources listed, including DisGeNET, DrugBank, UMLS, Orphanet, etc. in the paper https://www.nature.com/articles/s41597-023-01960-3 | Yes - Explicit versions and download dates provided for each dataset in the Methods/Data Records section | Partially - Tools like goatools, beautifulsoup, regex scripts, and vocabulary mappings are mentioned in the GitHub repo, but not all formal dependencies are listed | Yes - Each node contains node source; edges are annotated by type and origin | Yes - Duplicates and self-loops were removed during KG preprocessing and merging | Yes - Clear documentation on what triples were derived from which resource (e.g., drug–protein from DrugBank, phenotype–disease from HPO) | Yes - The paper documented schema of 30 edge types and their origin ontologies | Yes - Clinical and pharmacological text features are readable and interpretable (e.g., Mayo Clinic descriptions, DrugBank pharmacodynamics) | Yes - Uses Mondo, DrugBank, HPO, MeSH, Entrez Gene IDs, and UMLS CUIs, which are mappable and resolvable via external resources | Yes - Extensive paper + GitHub repo | Yes - Transformations like self-loop removal, duplicate dropping, phenotype-disease resolution, and mapping across ontologies are documented | Yes - Node and edge formats, and their standardized schema, are explained in the methodology and data files | No - No version tags (e.g., v1.0, v1.1) are mentioned or used on Dataverse or GitHub | No - GitHub Issues tab is not actively used for public feature requests or bug tracking | Yes - Maintained by Zitnik Lab at Harvard with lab contact and GitHub maintainers listed | No - Only one release version is available as of now (May 2022) | No - No archived prior versions or changelog indicating updates | Yes - Autism case study demonstrates disease concept resolution and clinical alignment | Yes - Compared to other KGs (e.g., SPOKE); benchmarks and references to prior systems included | Yes - Focused on disease-centric precision medicine with defined coverage: 17,080 diseases, 10 biological scales, 20 sources | Yes – Structure connectivity, edge density, text embedding-based grouping, and clinical relevance tested | Partially - Uses similarity thresholds (e.g., cosine ≥ 0.98 for disease grouping); no formal metrics like precision/recall provided | CC BY 4.0 |
| **HetioNet** | Yes - Can access paths, DWPCs, prediction probabilities, network support breakdowns for compound–disease pairs (via Neo4j Browser & guides) | Yes - fully hosted on a public Neo4j instance with Cypher queries, guides, tutorials https://neo4j.het.io/browser/ | Yes - Downloadable as JSON, Neo4j DB, TSV; also query online in Neo4j Browser; source code & intermediate datasets on GitHub, Zenodo, Figshare | Yes, the source code and scripts are public on hetio and GitHub linked in the paper https://github.com/elifesciences-publications/hetionet | Yes - Multiple export formats (JSON, Neo4j dump, TSV) | Yes - 29 sources documented; each node/edge carries source information in properties; full list with versions in paper https://elifesciences.org/articles/26726 | Yes - Versions noted: e.g., DrugBank v4.2, SIDER v4.1, LINCS L1000 (Oct 2015), Pathway Commons (with date) | Yes - Input ontologies and databases fully listed with versions; also intermediate resources described (e.g., STARGEO, PharmacotherapyDB | Yes - Node/edge properties include URLs, source, license, confidence scores (for applicable edges) | Yes - Merged redundant pathways; multiple studies for same edge consolidated; non-informative gene sets removed | Yes - Explicit per metaedge: e.g., binding affinities (≤1 mM), co-occurrence p-values (MEDLINE), gene interaction specifics | Yes - Clear metagraph with 11 node types & 24 metaedges; each with documented origin & justification | Yes - Uses standard biomedical IDs: Entrez, UMLS, MeSH, DO, Uberon | Yes - Entrez Gene, DOID, MeSH IDs, InChIKeys used for easy cross-referencing | Yes - Extensive: paper + Thinklab logs + GitHub issues + detailed guides | Yes - Explained pruning (e.g., filtering Uberon terms, merging pathways, restricting GO terms by size) | Yes - Metagraph is the explicit schema; node and edge types clearly defined | Partial - "v1.0" labeled, but no formal version history beyond initial | Partial - Thinklab (now static); issues can be filed on GitHub | Yes - Daniel Himmelstein and team, contactable via GitHub, Thinklab archives, paper | No - Only v1.0 publicly released so far | No - Early versions mentioned but no archived download versions listed | Yes - Nicotine dependence (bupropion), epilepsy predictions (acamprosate) | Yes - Compared to PREDICT, Guney et al., Cheng et al.; used baselines & permutation | Yes - Designed for systematic drug repurposing + broader knowledge integration | Yes - DWPC + AUROC + permutation + cross-validation + external test sets (DrugCentral, ClinicalTrials.gov) | Yes - Probability scores, cross-validated elastic net, path-level contribution breakdowns, AUROC | CC0 1.0 |
| **SPOKE** | yes - neighborhood explorer; limited to 10k nodes; API https://spoke.rbvi.ucsf.edu/swagger/ | yes - neighborhood explorer; limited to 10k nodes | yes | no | no | yes in supplemental and on website | yes; pulled each week | no | yes https://spoke.rbvi.ucsf.edu/docs/index.html | yes; sort of; genes and protiens are considered different | yes; some for instance specifies PPI for humans comes from STRING; human disease ontology provides disease associated gene | yes | yes no custom data transforms that are KG specific | yes uses Ensembl | yes edge and node types specified with source | yes; data is downloaded and integrated as is | yes uses Biolink when practical | no | no | yes email provided | yes updated weekly according to paper | no | no | no | yes; biomedical applications such as drug reuposing, disease prediction, embedding EHR data | no | no | CC BY 4.0 |
| **GNBR** | yes can access themes which relate to the path types and their score | no | no | no but can get code to process data | yes but its very weirdly done; separated into files by edge types | yes - but NLP so not many databases used | yes - pubtator annotations from april 30, 2016 | yes detailed dependancy path extraction; | yes each edge has a lot of information; NLP was used and the sentance that constructed something is listed | yes counted as a co occurance; feature not a bug used to inform network | yes - very limited | yes - NLP; had four interaction types and three node types | yes all information coming from publications outside of KG | yes uses UMLS IDs for rare diseases | yes specifies where disease and drug nodes come from | yes eliminated paths containing dependencies of type conj | yes self documented and reported with downloads | yes 7 versions available | no | no | yes but not updated since 2019 | yes and indicated which version was used in the paper | yes drug disease indications ; 2 case studies | yes; compared to gold standard drug indications; relationships from other databases for those gotten using NLP methods | yes drug repurposing | yes case studies; drug-disease path analysis; validation using DGIdb; used lit review to verify novel drug | yes performance on gold standard drug indications calculate AUROC; dendrograms of the themes | CC BY 4.0 |
| **NCATS Gard** | Yes – The KG enables extraction of disease mappings, disease profiles, and pathogenesis paths, as shown in the case studies (e.g., 3-hop paths from Wilson disease) | Yes – Public Neo4j instance available: https://disease.ncats.io | N | Partially – The in-house framework stitcher is mentioned and linked (https://github.com/ncats/stitcher), but full KG generation scripts for GARD KG are not public | Y (networkx pickle) | Y, integrates 34 distinct biomedical resources | ORPHADATA v4.0<br>Gene Ontology<br>Human Phenotype Ontology<br>GO annotations<br>Pathway Commons (PTC)<br>Pharos v3.8.0 | Partially – Tools like stitcher and OWL file parsing are mentioned, but not declared formally like a software repo would | Yes – each node/edge carries provenance via stitch keys (e.g., N_Name, I_CODE) | Partially – Mappings and harmonization discussed, but no explicit deduplication process for edges described | Yes – Object properties defined per source and integration method (e.g., has_phenotype, I_CODE, N_Name) | Yes – Meta-ontology schema is described with object properties and their meanings (Table 3) | Yes – Disease profiles, drug-disease associations, and harmonization rules demonstrate biomedical applicability | The paper explicitly mentions using "Unique Ingredient Identifier (UNII)," "MONDO ID," and "OMIM ID" for mappings, indicating that external identifiers are a core part of their entity resolution strategy | Yes – Detailed methods section describes data collection, mapping strategies, and harmonization | The paper notes that "data cleanup was performed," mentioning specific examples like restricting prefixes for OMIM IDs and Orphanet IDs. It also discusses the challenges of programmatic annotation, such as preventing the annotation of generic terms like "Disease" or "Syndrome" | The paper outlines a "meta-ontology" that serves as the schema for the knowledge graph. This includes a clear definition of primary classes and object properties used to structure the data; documented in tables 2–4 | No | No | Yes – Corresponding author: qian.zhu@nih.gov and NCATS/ NIH team affiliation listed | Yes - Suggests updates over time; indicates continually updated | No | Yes – Four detailed case studies: disease mapping, disease profiling, data harmonization, pathogenesis exploration | The paper compares its approach with other similar efforts, such as the semantic Diseasecard and the Monarch Initiative, placing its own work in the context of the broader field; but no explicit benchmarking | Yes – Focused on rare diseases, especially those curated in GARD, and integrating relevant biomedical data | Partially – Demonstrates utility via case studies but lacks formal quantitative evaluation (mapping stats in tables 6,7,10) | The paper's semi-automatic mapping process for FDA orphan designations includes a manual curation step where curators labeled mappings as "Done," "Approximate," or "Failed," providing a clear way to measure the confidence or accuracy of the mappings | CC BY 4.0 |
| **Monarch KG** | Yes, can access the mapping schema biolink model which provides sematics mappings and ontological relationships, can access downloadable graph dumps in SciGraph or KGX formats. | Yes, can access by Monarch's own webpage, API, Neo4j, downloadable Graph. | Yes, Monarch provides multiple ways to download the data such as KGX TSV, KGX JSON lines, Neo4j Dump, etc. | Yes, can access the source code for making the KG on Github. | Yes, KGX TSV, KGX JSON lines, Neo4j Dump, etc. | Yes, 33 heterogeneous data sources including HPOA, CTD, OMIM, Orphanet, WormBase, FlyBase, MGI, dictyBase, Xenbase, SGD, RGD, PomBase, ZFIN, NCBI, HGNC, Panther, BGeeDB, Reactome, STRING. | Yes, provides the version info of the sources used: https://data.monarchinitiative.org/monarch-kg-dev/latest/index.html | Yes, declared dependencies through its build infrastructure in the monarch-ingest and kg-hub Github repo. Specifically, Monarch uses Poetry for the dependencies management. | Yes, nodes and edges contain the most upstream source and knowledge provider information. | Yes, duplcate edge and node management is taken into consideration in the construction of the KG. | Yes, the triples created are captured in multiple stages of the KG ingest pipeline including the biolink preddicate mappings and the output. | Yes, Biolink model. All relationships are mapped to Biolink predicates. | Yes, Monarch KG is both human-readable and usable. Monarch provides its data in tablular formats that are easy to understand. Also, Monarch KG data and its Monarch initaitive webpage can be used in bioinformatics & biomedical analysis, machine learning, NLP, LLMs, visualization & exploration, etc. | Yes, supports resolvable and externally linked identifiders. Monarch KG uses curie-style IDs that follow identifier conventions and are mappable to URLs for resolution. Also, it includes a column called xref which contains a list of external identifiers. | Yes, Monarch made great effort to document its KG construction including data ingestion, transformation, schema usage, and KG merging. | Yes, Monarch documented data transforms including the excluded nodes and dangling edges. | Yes, Monarch uses a documented schema, Biolink Model, and its monarch-ingest pipeline for construction. | Yes | Yes, Monarch provides public tracker for feature requests, bug reports on it Github repo. | Yes, the contact is Monarch Initiative. Contact information is provided. | Yes, the KG is updated every month. | Yes, prior versions are accessible. And there is a dashoboard for QC showing what the changes are between any two versions. https://qc.monarchinitiative.org/#monarch?dataset=Development&kgVersion=2025-07-09 | Yes, Monarch KG is used in Exomiser which is a tool to annotate variants and has a ChatGPT plugin through RAG to support information search. It is also integrated into the GRAPE library for graph analysis and machine learning. | No | Yes, Monarch aims to harmonize the data across the fields (gene, disease, phenotype across species) to facilitate the discovery of disease mechanism and aid the disease diagnosis. | Yes, created Monarch Quality Control (QC) Dashboard for quality metrics that are specific to Monarch. In addition, it developed Phenotypic Inference Evaluation Framework (PhEval) to evaluate the analysis of its tool Exomiser. | Kind of but not explicit. The edge data contains columns like provided_by, publications, and has_evidence to measure the accuracy or confidence. | CC BY 4.0 |
| **clinical KG** | no | yes both | yes; neo4j, docker, csv, python library | yes well documented github https://github.com/MannLabs/CKG?tab=readme-ov-file | yes many different ways | yes in github with links as well as supplemental table S1 | no | yes in github and documents on how to download and use | yes edges have pubmed identifiers | yes; built in redundancy to assess overlap and disagreememnt of sources | no | no | yes use all external identifiers; ontolgies | yes used standardized identifiers for nodes not described with exsiting terminoloies or ontologies | yes well documented in the paper and a github is well done | no | yes custom; defined in paper figure 3A | yes three versions | yes; encouraged through github | yes mann labs | yes; newest is from 2022 | yes | yes; non-alcoholic fatty liver disease; cancer biomarkers; identify theraputic options for chemorefractory cases | yes compared their ability to identify cancer biomarkers with paper from 2018 | yes harmonization of proteomics with other omics data | no | no | C BY-NC 4.0 |
| **RTX-KG2** | Yes, RTX-KG2 is able to provide various biomedical information through queries. It is also used as backend to support ARAX's path reasoning and path ranking (https://github.com/RTXteam/RTX) . | Yes, can access by API query, and Neo4j. | Yes, multiple ways to access including downloadable versions, API (SmartAPI), web browser user interface (seems not currently working). | Yes, Github (https://github.com/RTXteam/RTX-KG2) | Yes, downloadable versions are available on Github | Yes, 70 sources Table 1 (UMLS, SemMedDB, ChEMBL, DrugBank, Reactome, SMPDB, and 64 additional knowledge sources). | Yes, it documents the versions of the upstream sources used (https://github.com/RTXteam/RTX-KG2/blob/master/docs/kg2-versions.md). | Yes, in the requirements file. | Yes, node's ID contains source information and edge contains primary knowledge source. | Yes, it provides a pre-canonicalized graph version (RTX-KG2pre, with semantically duplicated concepts) and a canonicalized version (RTX-KG2c, withthout semantically duplicated concepts) | Yes, in the final output KG, each edge includes the source that reated that triple. | Yes, it uses Biolink Model for as the schema standard for both nodes and edges. | Yes, it is used for other biological applications such as answering translational science questions, drug repositioning, identifying new therapeutic targets, and understanding drug mechanisms. | Yes, it uses resolvable IDs for the entities. | Yes, it has clear and step by step documentation on construction on its Github repo | Yes, in Appendix | Yes, Biolink model and extract-transform-load (ETL) approach for construction. | Yes, it is using semantic versioning (e.g., KG2.7.3) | Yes, provides public tracker for requests, bug reports on it Github repo. | Yes, it provides contact information of the KG2 Team. | Yes, once per month (mentioned in Discussion). | Yes, the prior versions are accessible (https://github.com/ncats/translator-lfs-artifacts/blob/main/README.md) with documented changes (https://github.com/RTXteam/RTX-KG2/blob/master/docs/kg2-versions.md). | Yes, it is currently being used by multiple Translator reasoning agents such as ARAX (Autonomous Relay Agent X). | Yes, it is compared to four other KGs (Hetionet, SPOKE, the SRI Reference Knowledge Graph, and ROBOKOP) | Yes, it is a part of NCATS Biomedical Data Translator project to support automated biomedical reasoning and question answering. It aims to create a semantically standardized, computable, and interoperable biomedical KG that supports translational reasoning and biomedical discovery. | Yes, it is not only evaluated with other KGs, but also evaluated on the tools that utilize it such as ARAX, mediKanren, BioThings Explorer, and ARAGORN. | Yes, the nodes and edges contain evidence, provenance, and other information for measuring accuracy and confidence. | CC BY 4.0 |
| **HRA-KG** | Yes, users can ask biological questions via queries (Basic Usage: https://cns-iu.github.io/hra-kg-supporting-information/#basic-usage, Advanced Usage: https://github.com/cns-iu/hra-kg-supporting-information/blob/main/README.md) | Yes, can access by SPARQL endpoint, HRA API, and RESTful-API. | Yes, dynamic queries through its SPARQL, RESTful-API, and UIs. | Yes, Github (https://github.com/hubmapconsortium/hra-kg) | Yes, available on Zenodo and HRA CDN | Yes, Table 2 | Yes, the type, name, and version of each DO is documented (https://lod.humanatlas.io/asct-b/). | Yes, in the hra-do-processor. | Yes, both nodes and edges have source information | Not explicit. As each digital object produces its own graph, the triples from different objects may have similar relationships, but they are stored in separate contexts. | Yes, the triples are generated by Digital Objects in in RDF format. | Yes, the KG is built based on RDF graph (CCF ontology) and standard ontologies to define relationship types between the entities. | Yes, the data is human readable and usable in various biological applications. | Yes, it uses resolvable IDs for the entity resolution such as UBERON:0013702 and ENSG00000175899. | Yes, it provides comprehensive documentation on its construction process (HRA KG Construction and Deployment). | Yes, it documents data transfroms such as data normalization and enrichment. | Yes, ASCT+B tables and Common Coordinate Framework (CCF) Ontology. | Yes, e.g., v2.3. | Not explicitly mentioned, but there are multiple Github Issues portals for users to post requests, reports bugs, and discussions. | Yes, contact information is available to reach out to the team (https://humanatlas.io/about#editorial-board) | Yes, about 2 time a year (https://github.com/hubmapconsortium/hra-kg/releases). | Yes, prior versions are accessible on Github (https://github.com/hubmapconsortium/hra-kg/releases). Changed are documented (https://humanatlas.io/release-notes/v2.3). | Yes, they have several use cases and applications (Usage Notes). | No, The HRA-KG is compared with other KGs only from technical perspective, such as the number of nodes, edges, and technologies used. | Yes, HRA-KG is focused on representing the healthy human body in semantics terms including anatomical structures, cell types, and biomarker and spatial terms such as 3D reference organs. | Yes, it has hra-ols used by OLS for validation (https://lod.humanatlas.io/collection/hra-ols/latest/) and has Validation report (https://github.com/hubmapconsortium/3d-hra-ref-object-validation) (Supplementary). | Not explicit, but it has HRAlit which links the HRA DOs to publications, experts, experimental datasets (Supplementary). | CC BY 4.0 |
| **Petagraph** | yes, portions of the graph that are open licensed are available from https://osf.io/6jtc9/files/osfstorage | yes, (ish), neo4j dump can be downloaded if you have a UMLS license key, and there are instructions to build a neo4j from sources in the github repository. Ish, because it's not specifically hosted, you need to host it yourself. | no, (ish) completed graph can only be downloaded as neo4j, components to build the graph can also be downloaded, but that doesn't quite feel like it makes a yes | yes, ETL of upstream sources lives in https://github.com/x-atlas-consortia/ubkg-etl/ | yes (with UMLS api key) | yes, methods section of paper https://www.nature.com/articles/s41597-024-04070-w#Sec2 | yes, ish, some source versions listed in https://github.com/TaylorResearchLab/Petagraph/tree/main/Scientific_Data_2024 | yes, ubkg etl has requirements.txt, petagraph has requirements-test.txt | Yes. They either come from existing ontologies or have one file per datasource with edges and nodes that are being added. | Yes - Yes. Bidirectional edges are only for Concept–Concept; other edges are unidirectional. Redundancies are minimized using binning and source normalization | Yes, methods of paper | yes custom schema defined in paper | Yes, csv files | Yes, concept CUI with codes corresponding to common external ID/s | Yes, well documented github | guidlines for formatting ontologies in user guide, method explain preprocessing steps | Start with ontologies and standards in the UBKG and add in omics data based on their paper defined in the schema | No, lists date last updated but no version | No | Yes, contributors listed https://osf.io/6jtc9/ | Yes, frequent small updates | recent acitivty documents date of all changes and person who made them but prior versions not accessible https://osf.io/6jtc9/ | Yes, several in paper | No | Yes integrating/analyzing multiomics datasets | link prediction tasks auROC, precision-recall, top tissues associated w a disease, and shortest path analysis of subgraphs | Yes. AUC-ROC, Precision-Recall curves, common neighbors vs random, structural metrics compared to randomized graphs | CC BY-NC-ND 4.0 |
| **GenomicKB** | Y, can submit an online query and download subgraphs that meet criteria | website with UI for creating queries, API supposely coming in v2, Neo4j dump available but not hosted by creators | Y, can use online query tool or download Neo4j dump | Y, https://github.com/tinalee-tech/KG-dataloader | Y, Neo4j dump | Y, supplementary note 1 | N | N | Y, tables 1 (nodes) & 2 (edges), query allows users to restrict by datasource | Considered but not de-duplicated. Users can restric querys to specific datasources and duplicate edges can be used for validation. EX: 'enhancer regulate gene' with restriction 'cell_line=GM12878' and 'data_source=EnhancerAtlas' -> 118,610 node pairs. A second query for 'variant overlap enhancer', 'enhancer regulate gene', and 'variant correlate_with gene' -> 16,871 enhancer-gene pairs from EnhancerAtlas can be validated by GTEx eQTLs | Y, table 2 | Y, table 2 | Y, they have a web interface (https://gkb.dcmb.med.umich.edu/) presenting the entities and edges in a human-readable format. It can be used to answer genomics-related questions and conduct biological analysis | Y, uses one external ID when possible and defines globaly unique IDs for entities without such as ChIP-seq peaks | Y, github and supplement | Y, sup Note 1 Data Collection and Processing | Uses custom schema not an existing one but does provide documentation | Y but still on version 1.0 despite multiple upates listed | N, FAQ says to email the team | Y, website has a contacts button with name/emal for both technical details and accessing the data/web server | During 2022-2023 yes but not since then | Clearly documented chages with dates but prior versions not available. "Paper states that they store all versions internally and can revert if unexpected changes are ever made to their graph but these are not accessible publicaly | The paper provides examples of specific queries that can be performed through the website that would normally require coding and integrating multiple data sources, but it is limited to knowledge extraction and does not show how this could be used to make novel biological discoveries | N | Y, coding free queries of human genomic information including both functional, strutural, epigenetic and other relationships | N | Kind of but not explicitly. Since this tool is all about data querying, confidence could be assesed by creating multiple queries like the example of using eQTLs to "validate" EnhancerAtlas relationships | CC BY-NC 4.0 |
| **ROBOKOP** | Y, can access subgraphs through query searches | Y: neo4j, ExEmPLAR (experimental tool for queries of Neo4j graphs), website with "question builder" feature, automat queries (cypher and TRAPI) | Y, can download or query searches in multiple ways | Y, https://github.com/RobokopU24/ORION (this is not the original GH from the paper but is now the one they are using) | Y, https://stars.renci.org/var/plater/graphs/RobokopKG/6fe13d850fdbf89c/ | Y, metadata contains all source info | Y metadata includes links and versions for all sources | Y, github lists docker requirements and graph spec file with example provided | Y, node and edges contain primary knowledge source attribute and, if relationship is inferred, any algorithms/methods used | not explicitly mentioned in paper or docs, but github for creating KG/metadata records have an "edge_normalization_version" so it appears they do | "one-hop" connection schema has summary of all edge types (node 1 type, node 2 type, relationship), and individual edges include the original data source, but sources per triple type were not explicitly stated/summarized | "one-hop" connection schema has summary of all edge types, and individual edges include the original data source, but sources per edge type were not explicitly stated | Yes, json files | Y, choose one external representative entity ID and map synonyms using existing equivalencies and specailized rules | Y, github repo for construction with reasonable documentation on how to use it in general, and metadata files include exact versions of functions that were used when creating their KG | not explicitly described but metadata contains all preprocessing steps per source with versions of each normalization strategies used | BioLink | Y, https://stars.renci.org/var/plater/graphs/RobokopKG/6fe13d850fdbf89c/RobokopKG.meta.json | Y, github | Contact form provided on website and weekly office hours over zoom | Yes - ROBOKOP was built 6 times in 2024 and 3 times so far this year. - as in this metadata file - https://stars.renci.org/var/plater/graphs/RobokopKG/6fe13d850fdbf89c/RobokopKG.meta.json | accessible for bulk download. No chnage log, but each comes with metadata that documents all source ingests (including versions) and preprocessing steps used for each source | Y, 4 ranging from simple linear queries to discovering mechanistic links | N | Yes - the scope is clearly defined because the paper focuses narrowly on presenting ROBOKOP KG and KGB as tools for federating biomedical sources into a usable KG, illustrating this through a few case examples. It does not try to claim to solve all KG problems, but sticks to integration via APIs and synonymization | N | Y, informative subgrahs are ranked according to number of supporting publications and literarure co-occurence of entity pairs | Open-Source MIT License |
| **EmBiology** | N | N | N | N | N | N | N | N | N | N | N | Y, changed yearly in their baseline updated but cannot see what schema is without paid account | N | Not explicitly stated but they use ML for entity recognition so probably | N | N, lots of this is based on NLP/text mining approaches but preprocessing of sources is not documented | N | N | N | Contact form (for sales team) | Y, cannot confirm but website states yearly update of terminology/rules, quarterly update of clinical trial data, weekly update of relationships extracted from publications | N | N | N | Not clearly stated but appears to be drug discovery for private companies | N | Confidence thresholds for relationships determined by number of supporting publications | Copyrighted to Elsevier |
| **DrugMechDB** | Y, paths that describe the mechanism of action for drug/disease indication pair | website, can view all paths or only those related to a specific drug or disease | Y Can download from zenodo or use website | Y: https://github.com/SuLab/DrugMechDB | Kind of on zenodo, but what they are calling a KG is a collection of entries for drug/disease pairs and the paths connecting them. Raw KG containing all edges between all entries does not appear to exist | table with node sources. Indications: DrugCentral ( according to GH, randomly sampled from availble drug-disease pairs, does not include all of them), paths: Mechanism of Action section from DrugBank, description section within Inxight Drugs, review articles, GeneOntology , UniProt, Reactome, and well-sources Wikipedia articles. No Primary experimental results are included. All are verified by curators | Date given for DrugCentral download, cannot find info for other sources | Y, there are requirements.txt and .yaml files in the KG's downloadable .zip file on Zenodo (https://zenodo.org/records/8139357) showing the dependencies. | each entry consist of 'graph', 'links', 'nodes', and 'reference' keys. Graph contains the drug/disease pair, links contains relationships informative to the indication, nodes contains information about all entities in link, and reference contains all the sources for the entire relationship but not specifically for individual nodes/edges | paths are manually curated | edges between informative nodes in a path do not contain orignal source infromation. Informative paths have citations for the entire path, not indivudal triples within them. | BioLink, but does not specify exactly what sources are used for each edge type | yes, json files | yes, all nodes come from existing sources | Not really. Everything is "manually curated" | N | BioLink | Y | GitHub Issues. Also includes a guide for how users can submit their own curated paths. | indirectly. website indicated the lab who owns it, but does not provide a point of contact | N (V1 released in 2019 with one update less than a year later, same with V2 in 2023, no other updates) | All versions availble, V2 has publication associated but detailed change logs for minor updates not available | N | They used an external KG to evaluate agreement between types of associations present in both graphs and calculate a p-value representing when the simulated percentage of matching was greater than or equal to the observed percentage | Yes, mechanisms of action by which a drug treats a disease | N | agreement with MechRepoNet | CC0 1.0 Universal |

# BioLink Node-Type Mappings

| Unstructured Label (Node-Type) | Biolink Class |
|---|---|
| Activity | activity |
| Agent | agent |
| Analytical Sample | material sample |
| Anatomy/Anatomical regions/Anatomical Structure | anatomical entity |
| Anatomical Entity | anatomical entity |
| Anatomy Cell Type | cell |
| Antibody | protein |
| Abtibody Panel | reagent targeted gene |
| Behavior | behavioral feature |
| Behavioral Feature | behavioral feature |
| Biological Activity | activity |
| Biological Entity | named thing |
| Biological Processes | biological process |
| Biological Sample | material sample |
| Biomarker | phenotypic feature |
| Cells | cell |
| Cell Lines | cell |
| Cell types | cell |
| Cellular Components | cellular component |
| Cell Summary | cell |
| Chemicals / Chemical Entity/ Drugs | chemical entity |
| Chemical Mixture | chemical entity |
| Chromosome | nucleic acid entity |
| Clinical Attribute | clinical attribute |
| Clinical Intervention | clinical intervention |

# BioLink Node-Type Mappings

| Unstructured Label (Node-Type) | Biolink Class |
|---|---|
| Clinical Variable | clinical attribute |
| Clinically Relevant Variable | clinical attribute |
| Code | information content entity |
| Cohort | population of individual organisms |
| Collision Summary | event |
| Complex | macromolecular complex |
| Compounds | chemical entity |
| Concept | information content entity |
| Corridor | geographic location |
| Dataset | dataset |
| Definition | information content entity |
| Devices | device |
| Diseases | disease |
| Domains | protein domain |
| Donor | individual organism |
| Enhancer | regulatory region |
| Environmental Features | environmental feature |
| Environmental Processes | environmental process |
| Enzymatic Activity | activity |
| Epigenomic Features | molecular modification |
| Event | event |
| Exon | nucleic acid entity |
| Experiment | study |
| Exposures | exposure |
| Extraction Site (Spatial Entity) | anatomical entity |

# BioLink Node-Type Mappings

| Unstructured Label (Node-Type) | Biolink Class |
|---|---|
| Food | food |
| FTU Illustration | information content entity |
| FTU Illustration Node | information content entity |
| Functional Regions | regulatory region |
| Genes | gene |
| Gene Family | gene family |
| Genomic Entity | genomic entity |
| Genotype | genotype |
| Geographical Location | geographic location |
| Gross Anatomical Structure | anatomical entity |
| GWAS Study | study |
| Information Content Entity | information content entity |
| Information Resource | information content entity |
| Insulator | regulatory region |
| Known Variant | sequence variant |
| Landmark | geographic location |
| Life Stage | life stage |
| Macro Molecular Complex | macromolecular complex |
| Material Sample | material sample |
| Metabolite | small molecule |
| MicroRNA | nucleic acid entity |
| Millitome Extraction Site | anatomical entity |
| Modification | molecular modification |
| Modified Protein | protein |
| Molecular Activity | activity |

# BioLink Node-Type Mappings

| Unstructured Label (Node-Type) | Biolink Class |
|---|---|
| Molecular Entity | named thing |
| Molecular Functions | activity |
| Molecular Mixture | chemical mixture |
| Named Thing | named thing |
| Non Coding RNA Product | nucleic acid entity |
| Nucleic Acid Entity | nucleic acid entity |
| Nutrients | small molecule |
| Organisms | organismal entity |
| Organism Taxon | organism taxon |
| Organismal Entity | organismal entity |
| Orthology | gene family |
| Pathological Process | pathological process |
| Pathways | pathway |
| Peptides | polypeptide |
| Perturbagens | drug |
| Pharmacologic Classes | chemical entity |
| Phenomenon | phenomenon |
| Phenotypes/ Effects/ Phenotypic Feature | phenotypic feature |
| Physical Entity | named thing |
| Physiological Processes | physiological process |
| Polypeptide | polypeptide |
| Population of Individual Organisms | population of individual organisms |
| Procedure | clinical intervention |
| Project | study |
| Promoter | regulatory region |

# BioLink Node-Type Mappings

| Unstructured Label (Node-Type) | Biolink Class |
|---|---|
| Proteins | protein |
| Protein Domains | protein domain |
| Protein Family | gene family |
| Protein Structure | protein |
| Publications | publication |
| Reactions | biological process |
| Reference Organ | anatomical entity |
| Reference Organ Part | anatomical entity |
| RNA Product | nucleic acid entity |
| Semantic | information content entity |
| Sequence Variant | sequence variant |
| Side Effects | phenotypic feature |
| Small Molecules | small molecule |
| Somatic Mutation | sequence variant |
| Subject | individual organism |
| Super Enhancer | regulatory region |
| Symptoms | phenotypic feature |
| Term | information content entity |
| Tissues | anatomical entity |
| Tssue Block (Sample) | material sample |
| Tissue Section (Sample) | material sample |
| Transcript | nucleic acid entity |
| Treatment | clinical intervention |
| User | agent |
| Variants | sequence variant |

# BioLink Node-Type Mappings

| Unstructured Label (Node-Type) | Biolink Class |
| --- | --- |
| Vitamins | small molecule |
| 3D Structures | protein domain |

# Node types

| | Total # KG per node type | RTX-KG2 | ROBOKOP | Clinical KG | SPOKE | HRA-KG | Monarch KG | GenomicKB | Bioteque | DrugMechDB | HetioNet | PrimeKG | NCATS Gard | Petagraph | PharmKG | GNBR | EmBiology* |
|---|---|---|---|---|---|---|---|---|---|---|---|---|---|---|---|---|---|
| **Total # of Node types** | | 53 | 36 | 34 | 21 | 19 | 17 | 14 | 13 | 13 | 12 | 10 | 7 | 5 | 4 | 3 | 0 |
| **Activity** | 1 | | ✓ | | | | | | | | | | | | | | |
| **Agent** | 1 | ✓ | | | | | | | | | | | | | | | |
| **Analytical Sample** | 1 | | | ✓ | | | | | | | | | | | | | |
| **Anatomy/Anatomical regions/Anatomical Structure** | 4 | | | | ✓ | ✓ | | | | | ✓ | ✓ | | | | | |
| **Anatomical Entity** | 3 | ✓ | ✓ | | | | ✓ | | | | | | | | | | |
| **Anatomy Cell Type** | 1 | | | | ✓ | | | | | | | | | | | | |
| **Antibody** | 1 | | | | | ✓ | | | | | | | | | | | |
| **Abtibody Panel** | 1 | | | | | ✓ | | | | | | | | | | | |
| **Behavior** | 1 | ✓ | | | | | | | | | | | | | | | |
| **Behavioral Feature** | 1 | ✓ | | | | | | | | | | | | | | | |
| **Biological Activity** | 1 | ✓ | | | | | | | | | | | | | | | |
| **Biological Entity** | 1 | ✓ | | | | | | | | | | | | | | | |
| **Biological Processes** | 8 | ✓ | ✓ | ✓ | ✓ | | ✓ | | | ✓ | ✓ | ✓ | | | | | |
| **Biological Sample** | 1 | | | ✓ | | | | | | | | | | | | | |
| **Biomarker** | 1 | | | | | ✓ | | | | | | | | | | | |
| **Cells** | 6 | ✓ | ✓ | | | | ✓ | | ✓ | ✓ | | | ✓ | | | | |
| **Cell Lines** | 2 | ✓ | | | | | | ✓ | | | | | | | | | |
| **Cell types** | 2 | | | | ✓ | ✓ | | | | | | | | | | | |

# Node types

| | Total # KG per node type | RTX-KG2 | ROBOKOP | Clinical KG | SPOKE | HRA-KG | Monarch KG | GenomicKB | Bioteque | DrugMechDB | HetioNet | PrimeKG | NCATS Gard | Petagraph | PharmKG | GNBR | EmBiology* |
|---|---|---|---|---|---|---|---|---|---|---|---|---|---|---|---|---|---|
| **Total # of Node types** | | 53 | 36 | 34 | 21 | 19 | 17 | 14 | 13 | 13 | 12 | 10 | 7 | 5 | 4 | 3 | 0 |
| **Cellular Components** | 9 | ✓ | ✓ | ✓ | ✓ | | ✓ | | ✓ | ✓ | ✓ | ✓ | | | | | |
| **Cell Summary** | 1 | | | | | ✓ | | | | | | | | | | | |
| **Chemicals / Chemical Entity/ Drugs** | 11 | ✓ | ✓ | ✓ | | | ✓ | | ✓ | ✓ | ✓ | ✓ | ✓ | | ✓ | ✓ | |
| **Chemical Mixture** | 1 | ✓ | | | | | | | | | | | | | | | |
| **Chromosome** | 1 | | | ✓ | | | | | | | | | | | | | |
| **Clinical Attribute** | 1 | | | ✓ | | | | | | | | | | | | | |
| **Clinical Intervention** | 2 | ✓ | ✓ | | | | | | | | | | | | | | |
| **Clinical Variable** | 1 | | | ✓ | | | | | | | | | | | | | |
| **Clinically Relevant Variable** | 1 | | | ✓ | | | | | | | | | | | | | |
| **Code** | 1 | | | | | | | | | | | | | ✓ | | | |
| **Cohort** | 1 | ✓ | | | | | | | | | | | | | | | |
| **Collision Summary** | 2 | | ✓ | | | ✓ | | | | | | | | | | | |
| **Complex** | 1 | | | ✓ | | | | | | | | | | | | | |
| **Compounds** | 3 | | | | ✓ | | | | ✓ | | ✓ | | | | | | |
| **Concept** | 1 | | | | | | | | | | | | | ✓ | | | |
| **Corridor** | 1 | | | | | ✓ | | | | | | | | | | | |
| **Dataset** | 1 | | | | | ✓ | | | | | | | | | | | |
| **Definition** | 1 | | | | | | | | | | | | | ✓ | | | |

# Node types

| | Total # KG per node type | RTX-KG2 | ROBOKOP | Clinical KG | SPOKE | HRA-KG | Monarch KG | GenomicKB | Bioteque | DrugMechDB | HetioNet | PrimeKG | NCATS Gard | Petagraph | PharmKG | GNBR | EmBiology* | |
|---|---|---|---|---|---|---|---|---|---|---|---|---|---|---|---|---|---|---|
| **Total # of Node types** | | 53 | 36 | 34 | 21 | 19 | 17 | 14 | 13 | 13 | 12 | 10 | 7 | 5 | 4 | 3 | 0 | |
| **Devices** | 1 | ✓ | | | | | | | | | | | | | | | | |
| **Diseases** | 12 | ✓ | ✓ | ✓ | ✓ | | ✓ | | ✓ | ✓ | ✓ | ✓ | ✓ | | ✓ | ✓ | | |
| **Domains** | 2 | | ✓ | | | | | | ✓ | | | | | | | | | |
| **Donor** | 1 | | | | | ✓ | | | | | | | | | | | | |
| **Enhancer** | 1 | | | | | | | ✓ | | | | | | | | | | |
| **Environmental Features** | 1 | ✓ | | | | | | | | | | | | | | | | |
| **Environmental Processes** | 1 | ✓ | | | | | | | | | | | | | | | | |
| **Enzymatic Activity** | 1 | | | | ✓ | | | | | | | | | | | | | |
| **Epigenomic Features** | 1 | | | | | | | ✓ | | | | | | | | | | |
| **Event** | 1 | ✓ | | | | | | | | | | | | | | | | |
| **Exon** | 2 | ✓ | | | | | | ✓ | | | | | | | | | | |
| **Experiment** | 1 | | | ✓ | | | | | | | | | | | | | | |
| **Exposures** | 1 | | | | | | | | | | | ✓ | | | | | | |
| **Extraction Site (Spatial Entity)** | 1 | | | | | ✓ | | | | | | | | | | | | |
| **Food** | 3 | ✓ | | ✓ | ✓ | | | | | | | | | | | | | |
| **FTU Illustration** | 1 | | | | | ✓ | | | | | | | | | | | | 8 |
| **FTU Illustration Node** | 1 | | | | | ✓ | | | | | | | | | | | | |
| **Functional Regions** | 1 | | | ✓ | | | | | | | | | | | | | | |

# Node types

| | Total # KG per node type | RTX-KG2 | ROBOKOP | Clinical KG | SPOKE | HRA-KG | Monarch KG | GenomicKB | Bioteque | DrugMechDB | HetioNet | PrimeKG | NCATS Gard | Petagraph | PharmKG | GNBR | EmBiology* |
|---|---|---|---|---|---|---|---|---|---|---|---|---|---|---|---|---|---|
| **Total # of Node types** | | 53 | 36 | 34 | 21 | 19 | 17 | 14 | 13 | 13 | 12 | 10 | 7 | 5 | 4 | 3 | 0 |
| **Genes** | 11 | ✓ | | ✓ | ✓ | | ✓ | ✓ | ✓ | | ✓ | ✓ | ✓ | | ✓ | ✓ | |
| **Gene Family** | 3 | ✓ | ✓ | | | | | | | ✓ | | | | | | | |
| **Genomic Entity** | 1 | | ✓ | | | | | | | | | | | | | | |
| **Genotype** | 2 | | ✓ | | | | ✓ | | | | | | | | | | |
| **Geographical Location** | 2 | ✓ | ✓ | | | | | | | | | | | | | | |
| **Gross Anatomical Structure** | 2 | ✓ | | | | | | | | ✓ | | | | | | | |
| **GWAS Study** | 3 | | ✓ | ✓ | | | | ✓ | | | | | | | | | |
| **Information Content Entity** | 2 | ✓ | ✓ | | | | | | | | | | | | | | |
| **Information Resource** | 1 | ✓ | | | | | | | | | | | | | | | |
| **Insulator** | 2 | | ✓ | | | | | ✓ | | | | | | | | | |
| **Known Variant** | 1 | | | ✓ | | | | | | | | | | | | | |
| **Landmark** | 1 | | | | | ✓ | | | | | | | | | | | |
| **Life Stage** | 2 | ✓ | | | | | ✓ | | | | | | | | | | |
| **Macro Molecular Complex** | 1 | | | | | | | | | ✓ | | | | | | | |
| **Material Sample** | 1 | ✓ | | | | | | | | | | | | | | | |
| **Metabolite** | 1 | | | ✓ | | | | | | | | | | | | | |
| **MicroRNA** | 1 | ✓ | | | | | | | | | | | | | | | |
| **Millitome Extraction Site** | 1 | | | | | ✓ | | | | | | | | | | | |

# Node types

| | Total # KG per node type | RTX-KG2 | ROBOKOP | Clinical KG | SPOKE | HRA-KG | Monarch KG | GenomicKB | Bioteque | DrugMechDB | HetioNet | PrimeKG | NCATS Gard | Petagraph | PharmKG | GNBR | EmBiology* |
|---|---|---|---|---|---|---|---|---|---|---|---|---|---|---|---|---|---|
| **Total # of Node types** | | 53 | 36 | 34 | 21 | 19 | 17 | 14 | 13 | 13 | 12 | 10 | 7 | 5 | 4 | 3 | 0 |
| **Modification** | 1 | | | ✓ | | | | | | | | | | | | | |
| **Modified Protein** | 1 | | | ✓ | | | | | | | | | | | | | |
| **Molecular Activity** | 3 | ✓ | | | | | ✓ | | | ✓ | | | | | | | |
| **Molecular Entity** | 2 | ✓ | | | | | ✓ | | | | | | | | | | |
| **Molecular Functions** | 6 | | ✓ | ✓ | ✓ | | | | ✓ | | ✓ | ✓ | | | | | |
| **Molecular Mixture** | 1 | | ✓ | | | | | | | | | | | | | | |
| **Named Thing** | 3 | ✓ | ✓ | | | | ✓ | | | | | | | | | | |
| **Non Coding RNA Product** | 2 | ✓ | | | | | | ✓ | | | | | | | | | |
| **Nucleic Acid Entity** | 2 | ✓ | ✓ | | | | | | | | | | | | | | |
| **Nutrients** | 2 | | ✓ | | ✓ | | | | | | | | | | | | |
| **Organisms** | 2 | ✓ | | | ✓ | | | | | | | | | | | | |
| **Organism Taxon** | 3 | ✓ | | | | | ✓ | | | ✓ | | | | | | | |
| **Organismal Entity** | 1 | ✓ | | | | | | | | | | | | | | | |
| **Pathological Process** | 1 | | ✓ | | | | | | | | | | | | | | |
| **Pathways** | 9 | ✓ | ✓ | ✓ | ✓ | | ✓ | | ✓ | ✓ | ✓ | ✓ | | | | | |
| **Peptides** | 1 | | | ✓ | | | | | | | | | | | | | |
| **Perturbagens** | 1 | | | | | | | | ✓ | | | | | | | | |
| **Pharmacologic Classes** | 4 | | ✓ | | ✓ | | | | ✓ | | ✓ | | | | | | |

# Node types

| | Total # KG per node type | RTX-KG2 | ROBOKOP | Clinical KG | SPOKE | HRA-KG | Monarch KG | GenomicKB | Bioteque | DrugMechDB | HetioNet | PrimeKG | NCATS Gard | Petagraph | PharmKG | GNBR | EmBiology* |
|---|---|---|---|---|---|---|---|---|---|---|---|---|---|---|---|---|---|
| **Total # of Node types** | | 53 | 36 | 34 | 21 | 19 | 17 | 14 | 13 | 13 | 12 | 10 | 7 | 5 | 4 | 3 | 0 |
| **Phenomenon** | 2 | ✓ | ✓ | | | | | | | | | | | | | | |
| **Phenotypes/ Effects/ Phenotypic Feature** | 6 | ✓ | | ✓ | | | ✓ | | | ✓ | | ✓ | ✓ | | | | |
| **Physical Entity** | 1 | ✓ | | | | | | | | | | | | | | | |
| **Physiological Processes** | 1 | ✓ | | | | | | | | | | | | | | | |
| **Polypeptide** | 2 | ✓ | ✓ | | | | | | | | | | | | | | |
| **Population of Individual Organisms** | 2 | ✓ | ✓ | | | | | | | | | | | | | | |
| **Procedure** | 2 | ✓ | ✓ | | | | | | | | | | | | | | |
| **Project** | 2 | | ✓ | ✓ | | | | | | | | | | | | | |
| **Promoter** | 1 | | | | | | | ✓ | | | | | | | | | |
| **Proteins** | 10 | ✓ | ✓ | ✓ | ✓ | | ✓ | ✓ | ✓ | ✓ | | | ✓ | | ✓ | | |
| **Protein Domains** | 2 | | ✓ | | ✓ | | | | | | | | | | | | |
| **Protein Family** | 1 | | | | ✓ | | | | | | | | | | | | |
| **Protein Structure** | 1 | | | ✓ | | | | | | | | | | | | | |
| **Publications** | 3 | ✓ | ✓ | ✓ | | | | | | | | | | | | | |
| **Reactions** | 1 | | | | ✓ | | | | | | | | | | | | |
| **Reference Organ** | 1 | | | | | ✓ | | | | | | | | | | | |
| **Reference Organ Part** | 1 | | | | | ✓ | | | | | | | | | | | |
| **RNA Product** | 2 | ✓ | ✓ | | | | | | | | | | | | | | |

# Node types

| | Total # KG per node type | RTX-KG2 | ROBOKOP | Clinical KG | SPOKE | HRA-KG | Monarch KG | GenomicKB | Bioteque | DrugMechDB | HetioNet | PrimeKG | NCATS Gard | Petagraph | PharmKG | GNBR | EmBiology* |
|---|---|---|---|---|---|---|---|---|---|---|---|---|---|---|---|---|---|
| **Total # of Node types** | | 53 | 36 | 34 | 21 | 19 | 17 | 14 | 13 | 13 | 12 | 10 | 7 | 5 | 4 | 3 | 0 |
| **Semantic** | 1 | | | | | | | | | | | | | ✓ | | | |
| **Sequence Variant** | 1 | | | | | | ✓ | | | | | | | | | | |
| **Side Effects** | 2 | | | | ✓ | | | | | | ✓ | | | | | | |
| **Small Molecules** | 1 | ✓ | | | | | | | | | | | | | | | |
| **Somatic Mutation** | 1 | | | ✓ | | | | | | | | | | | | | |
| **Subject** | 1 | | | ✓ | | | | | | | | | | | | | |
| **Super Enhancer** | 0 | | | | | | | | | | | | | | | | |
| **Symptoms** | 3 | | ✓ | | ✓ | | | | | | ✓ | | | | | | |
| **Term** | 1 | | | | | | | | | | | | | ✓ | | | |
| **Tissues** | 4 | | | ✓ | | | | ✓ | ✓ | | | | ✓ | | | | |
| **Tssue Block (Sample)** | 1 | | | | | ✓ | | | | | | | | | | | |
| **Tissue Section (Sample)** | 1 | | | | | ✓ | | | | | | | | | | | |
| **Transcript** | 3 | ✓ | | ✓ | | | | ✓ | | | | | | | | | |
| **Treatment** | 1 | ✓ | | | | | | | | | | | | | | | |
| **User** | 1 | | | ✓ | | | | | | | | | | | | | |
| **Variants** | 2 | | | ✓ | | | | ✓ | | | | | | | | | |
| **Vitamins** | 2 | ✓ | ✓ | | | | | | | | | | | | | | |
| **3D Structures** | 1 | | | | | | | ✓ | | | | | | | | | |

# KG SOURCES

| | Total # KG per source | Monarch KG | RTX-KG2 | ROBOKOP | Bioteque | SPOKE | GenomicKB | HetioNet | clinical KG | Petagraph | DrugMechDB | PrimeKG | NCATS Gard | HRA-KG | EmBiology | PharmKG | GNBR |
|---|---|---|---|---|---|---|---|---|---|---|---|---|---|---|---|---|---|
| **Total # of Sources** | | 87 | 63 | 50 | 40 | 47 | 23 | 29 | 33 | 26 | 19 | 14 | 20 | 10 | 6 | 7 | 1 |
| **Alliance of Genome Resources** | 1 | ✓ | | | | | | | | | | | | | | | |
| **ALZHEIMERS-UNIVERSITY-OF-TORONTO** | 1 | ✓ | | | | | | | | | | | | | | | |
| **AGBASE** | 0 | | | | | | | | | | | | | | | | |
| **ARGKB** | 1 | | | ✓ | | | | | | | | | | | | | |
| **ARUK-UCL** | 1 | ✓ | | | | | | | | | | | | | | | |
| **Anatomical Therapeutic Chemical Classification System (ATC)** | 1 | | ✓ | | | | | | | | | | | | | | |
| **ASP2019** | 1 | | | | | | | | | ✓ | | | | | | | |
| **Basic Formal Ontology (BFO)** | 2 | ✓ | ✓ | | | | | | | | | | | | | | |
| **Bgee** | 5 | ✓ | | ✓ | | ✓ | | ✓ | | | | ✓ | | | | | |
| **BHF-UCL** | 0 | | | | | | | | | | | | | | | | |
| **BindingDB** | 3 | | | ✓ | | ✓ | | ✓ | | | | | | | | | |
| **BioGRID** | 2 | ✓ | | | | | | | | | | | | | ✓ | | |
| **Biolink** | 3 | | ✓ | ✓ | | | | | | | ✓ | | | | | | |
| **Biological Spatial Ontology (BSPO)** | 1 | | ✓ | | | | | | | | | | | | | | |
| **Brenda Tissue Ontology (BTO)** | 3 | | | | ✓ | | ✓ | | ✓ | | | | | | | | |
| **CACAO** | 1 | ✓ | | | | | | | | | | | | | | | |
| **CAFA** | 1 | ✓ | | | | | | | | | | | | | | | |
| **Cancer Cell Line Encyclopedia (CCLE)** | 1 | | | | ✓ | | | | | | | | | | | | |
| **Cancer Genome Interpreter** | 1 | | | | | | | | ✓ | | | | | | | | |

# KG SOURCES

| | Total # KG per source | Monarch KG | RTX-KG2 | ROBOKOP | Bioteque | SPOKE | GenomicKB | HetioNet | clinical KG | Petagraph | DrugMechDB | PrimeKG | NCATS Gard | HRA-KG | EmBiology | PharmKG | GNBR |
|---|---|---|---|---|---|---|---|---|---|---|---|---|---|---|---|---|---|
| **Catalogue of Semantic Mutations in Cancer (COSMIC)** | 1 | | | | ✓ | | | | | | | | | | | | |
| **CCIDB** | 1 | | | ✓ | | | | | | | | | | | | | |
| **Cell Miner** | 1 | | | | ✓ | | | | | | | | | | | | |
| **Cell Ontology (CL)** | 7 | ✓ | ✓ | | | ✓ | ✓ | | | | ✓ | | ✓ | ✓ | | | |
| **Cellosaurus** | 1 | | | | ✓ | | | | | | | | | | | | |
| **Cell Line ontology (CLO)** | 1 | | | | | | | | | | | | ✓ | | | | |
| **ChEBI** | 6 | ✓ | ✓ | ✓ | ✓ | | | | | | ✓ | | ✓ | | | | |
| **ChEMBL** | 3 | | ✓ | ✓ | | ✓ | | | | | | | | | | | |
| **Chemical Checker** | 1 | | | | ✓ | | | | | | | | | | | | |
| **CIVic** | 1 | | | | | ✓ | | | | | | | | | | | |
| **ClinicalTrials.gov** | 2 | | | | | ✓ | | | | | | | | | ✓ | | |
| **ClinGen** | 1 | ✓ | | | | | | | | | | | | | | | |
| **ClinVar** | 2 | ✓ | | | | | | | | ✓ | | | | | | | |
| **CLUE** | 1 | | | | ✓ | | | | | | | | | | | | |
| **CoexpressDB** | 1 | | | | ✓ | | | | | | | | | | | | |
| **Common Coordinate Framework Ontology (CCF)** | 1 | | | | | | | | | | | | | ✓ | | | |
| **Comparative Toxicogenomics Database** | 0 | | | | | | | | | | | | | | | | |
| **Compartments** | 1 | | | | ✓ | | | | | | | | | | | | |
| **ComplexPortal** | 1 | ✓ | | | | | | | | | | | | | | | |
| **Connectivity Map (CMAP)** | 1 | | | | | | | | | ✓ | | | | | | | |

# KG SOURCES

| | Total # KG per source | Monarch KG | RTX-KG2 | ROBOKOP | Bioteque | SPOKE | GenomicKB | HetioNet | clinical KG | Petagraph | DrugMechDB | PrimeKG | NCATS Gard | HRA-KG | EmBiology | PharmKG | GNBR |
|---|---|---|---|---|---|---|---|---|---|---|---|---|---|---|---|---|---|
| **CORUM** | 2 | | | | ✓ | | | | ✓ | | | | | | | | |
| **CREEDS** | 1 | | | | ✓ | | | | | | | | | | | | |
| **CTD** | 4 | ✓ | | ✓ | ✓ | | | | | | | ✓ | | | | | |
| **Depmap** | 1 | | | | ✓ | | | | | | | | | | | | |
| **dbSNP** | 1 | | | | | | ✓ | | | | | | | | | | |
| **dbSuper** | 1 | | | | | | ✓ | | | | | | | | | | |
| **dbVar** | 1 | | | | | | ✓ | | | | | | | | | | |
| **DDPHENO** | 1 | ✓ | | | | | | | | | | | | | | | |
| **DDANAT** | 1 | ✓ | | | | | | | | | | | | | | | |
| **DFLAT** | 1 | ✓ | | | | | | | | | | | | | | | |
| **DGIdb** | 1 | | ✓ | | | | | | | | | | | | | | |
| **DGV** | 1 | | | | | | ✓ | | | | | | | | | | |
| **DIBU** | 1 | ✓ | | | | | | | | | | | | | | | |
| **Dictyostelium discoideum anatomy (DDANAT)** | 1 | | ✓ | | | | | | | | | | | | | | |
| **DictyBase** | 1 | ✓ | | | | | | | | | | | | | | | |
| **DISEASES** | 4 | | | ✓ | | ✓ | | ✓ | ✓ | | | | | | | | |
| **DisGenNET** | 7 | | ✓ | ✓ | ✓ | ✓ | | ✓ | ✓ | | | ✓ | | | | | |
| **DisPROT** | 1 | ✓ | | | | | | | | | | | | | | | |
| **DGIdb** | 1 | | | | | | | | ✓ | | | | | | | | |
| **DO** | 5 | | ✓ | | | ✓ | | ✓ | ✓ | | | ✓ | | | | | |

# KG SOURCES

| | Total # KG per source | Monarch KG | RTX-KG2 | ROBOKOP | Bioteque | SPOKE | GenomicKB | HetioNet | clinical KG | Petagraph | DrugMechDB | PrimeKG | NCATS Gard | HRA-KG | EmBiology | PharmKG | GNBR |
|---|---|---|---|---|---|---|---|---|---|---|---|---|---|---|---|---|---|
| DOAF | 2 | | | | | ✓ | | ✓ | | | | | | | | | |
| DoRoth EA | 1 | | | | ✓ | | | | | | | | | | | | |
| DrugBank | 9 | | ✓ | ✓ | ✓ | ✓ | | ✓ | ✓ | | ✓ | | | | ✓ | ✓ | |
| DrugCentral | 7 | | ✓ | ✓ | ✓ | ✓ | | ✓ | | | ✓ | ✓ | | | | | |
| DrugMechDB | 1 | | | ✓ | | | | | | | | | | | | | |
| Drug repurposing Hub | 1 | | | | ✓ | | | | | | | | | | | | |
| ehrlink | 2 | | | | | ✓ | | ✓ | | | | | | | | | |
| EMAPA(mouse developmental anatomy ontology) | 1 | ✓ | | | | | | | | | | | | | | | |
| ENCODE | 1 | | | | | | ✓ | | | | | | | | | | |
| ENdb | 1 | | | | | | ✓ | | | | | | | | | | |
| EnhancerAtlas | 1 | | | | | | ✓ | | | | | | | | | | |
| Entrez Gene | 2 | | | | | ✓ | | ✓ | | | | | | | | | |
| Ensembl | 3 | ✓ | ✓ | | | | ✓ | | | | | | | | | | |
| eRAM (encyclopedia of rare disease annotations for precision medicine) | 1 | | | ✓ | | | | | | | | | | | | | |
| ESPD | 1 | | | ✓ | | | | | | | | | | | | | |
| Eukaryotic Promoter Database (EPD) | 1 | | | | | | ✓ | | | | | | | | | | |
| Evolutionary Rate Covariation | 2 | | | | | ✓ | | ✓ | | | | | | | | | |
| Experimental Factor Ontology (EFO) | 2 | | ✓ | | | | | | ✓ | | | | | | | | |
| FAERS | 1 | | | ✓ | | | | | | | | | | | | | |
| FANTOM5 | 1 | | | | | | ✓ | | | | | | | | | | |

# KG SOURCES

| | Total # KG per source | Monarch KG | RTX-KG2 | ROBOKOP | Bioteque | SPOKE | GenomicKB | HetioNet | clinical KG | Petagraph | DrugMechDB | PrimeKG | NCATS Gard | HRA-KG | EmBiology | PharmKG | GNBR |
|---|---|---|---|---|---|---|---|---|---|---|---|---|---|---|---|---|---|
| **FBBT** | 1 | ✓ | | | | | | | | | | | | | | | |
| **FBCV** | 1 | ✓ | | | | | | | | | | | | | | | |
| **FBDV** | 1 | ✓ | | | | | | | | | | | | | | | |
| **FDA Orphan Drug Designations Data** | 1 | | | | | | | | | | | | ✓ | | | | |
| **FlyBase** | 2 | ✓ | | ✓ | | | | | | | | | | | | | |
| **FooDB** | 2 | | | | | ✓ | | | ✓ | | | | | | | | |
| **Food Ontology** | 1 | | ✓ | | | | | | | | | | | | | | |
| **Foundational Model of Anatomy (FMA)** | 2 | | ✓ | | | | | | | | | | | ✓ | | | |
| **FYPO** | 1 | ✓ | | | | | | | | | | | | | | | |
| **GARD** | 1 | | | | | | | | | | | | ✓ | | | | |
| **GDB** | 1 | ✓ | | | | | | | | | | | | | | | |
| **GDSC** | 2 | | | | ✓ | ✓ | | | | | | | | | | | |
| **GHR** | 1 | | | | | | | | | | | | ✓ | | | | |
| **GNBR** | 0 | | | | | | | | | | | | | | | | |
| **GO** | 10 | ✓ | ✓ | ✓ | ✓ | ✓ | ✓ | ✓ | ✓ | | ✓ | ✓ | | | | | |
| **GO-Central** | 1 | ✓ | | | | | | | | | | | | | | | |
| **GO consortium** | 1 | ✓ | | | | | | | | | | | | | | | |
| **GENCODE** | 1 | | | | | | | | | ✓ | | | | | | | |
| **GENOMICS** | 1 | | | | | | | | | ✓ | | | | | | | |
| **Genomic Epidemiology Ontology (GENEPIO)** | 0 | | | | | | | | | | | | | | | | |

# KG SOURCES

| | Total # KG per source | Monarch KG | RTX-KG2 | ROBOKOP | Bioteque | SPOKE | GenomicKB | HetioNet | clinical KG | Petagraph | DrugMechDB | PrimeKG | NCATS Gard | HRA-KG | EmBiology | PharmKG | GNBR |
|---|---|---|---|---|---|---|---|---|---|---|---|---|---|---|---|---|---|
| **Gene Ontology, with external relationships (GO-Plus)** | 1 | | ✓ | | | | | | | | | | | | | | |
| **GOC-OWL** | 1 | ✓ | | | | | | | | | | | | | | | |
| **GO-NOCTUA** | 1 | ✓ | | | | | | | | | | | | | | | |
| **GlyGen** | 1 | | | | | | | | | ✓ | | | | | | | |
| **GTEx** | 3 | | | ✓ | | | ✓ | | | ✓ | | | | | | | |
| **Guide to PHARMACOLOGY (GtoPdb)** | 1 | | | ✓ | | | | | | | | | | | | | |
| **GWAS Catalog** | 5 | | | ✓ | | ✓ | ✓ | ✓ | ✓ | | | | | | | | |
| **Harmonizome** | 1 | | | | ✓ | | | | | | | | | | | | |
| **Health Level Seven Version 3.0 (HL7V3.0)** | 1 | | ✓ | | | | | | | | | | | | | | |
| **Hetio-dag** | 3 | | | ✓ | | ✓ | | ✓ | | | | | | | | | |
| **Hetionet** | 1 | | | | ✓ | | | | | | | | | | | | |
| **Healthcare Common Procedure Coding System (HCPCS)** | 1 | | ✓ | | | | | | | | | | | | | | |
| **HPOMP** | 0 | | | | | | | | | | | | | | | | |
| **Homo Sapiens Chromosomal Location Ontology (HSCLO38)** | 1 | | | | | | | | | ✓ | | | | | | | |
| **HPO** | 8 | ✓ | ✓ | ✓ | | | | | ✓ | ✓ | ✓ | ✓ | ✓ | | | | |
| **HuBMAP Research Attributes Value Set (HRAVS)** | 2 | | | | | | | | | ✓ | | | | ✓ | | | |
| **Human Developmental Anatomy, abstract** | 1 | | ✓ | | | | | | | | | | | | | | |
| **Human gene nomenclature committee (HGNC)** | 5 | ✓ | ✓ | ✓ | | | | | | ✓ | | | | ✓ | | | |
| **HGNC-UCL** | 1 | ✓ | | | | | | | | | | | | | | | |
| **Human Interactome Database** | 2 | | | | | ✓ | | ✓ | | | | | | | | | |

# KG SOURCES

| | Total # KG per source | Monarch KG | RTX-KG2 | ROBOKOP | Bioteque | SPOKE | GenomicKB | HetioNet | clinical KG | Petagraph | DrugMechDB | PrimeKG | NCATS Gard | HRA-KG | EmBiology | PharmKG | GNBR |
|---|---|---|---|---|---|---|---|---|---|---|---|---|---|---|---|---|---|
| **Human metabolite Database (HMDB)** | 3 | | ✓ | ✓ | | | | | ✓ | | | | | | | | |
| **Human-to-mouse ortholog mappings (HCOP)** | 1 | | | | | | | | | ✓ | | | | | | | |
| **Human-to-mouse phenotype mappings (HPOMP)** | 1 | | | | | | | | | ✓ | | | | | | | |
| **HSAPDV** | 1 | ✓ | | | | | | | | | | | | | | | |
| **Human-to-rat ENSEMBL mappings (RATHCOP)** | 1 | | | | | | ✓ | | | | | | | | | | |
| **HumanNet** | 1 | | | | | | | | | | | | | | | ✓ | |
| **Human Protien Atlas** | 4 | ✓ | | | ✓ | ✓ | | | ✓ | | | | | | | | |
| **HuRI** | 1 | | | | ✓ | | | | | | | | | | | | |
| **ICD-10 procedure coding system (ICD10PCS)** | 1 | | ✓ | | | | | | | | | | | | | | |
| **ICD-9, clinical modification (ICD9CM)** | 1 | | ✓ | | | | | | | | | | | | | | |
| **Incomplete Interactome** | 2 | | | | | ✓ | | ✓ | | | | | | | | | |
| **Inxight Drugs** | 2 | | | | | | | | | | ✓ | | ✓ | | | | |
| **IntAct** | 4 | | ✓ | ✓ | ✓ | | | | ✓ | | | | | | | | |
| **Interaction Network Ontology (INO)** | 1 | | ✓ | | | | | | | | | | | | | | |
| **InterPro** | 4 | ✓ | | | ✓ | ✓ | | | | | ✓ | | | | | | |
| **INTACT** | 1 | ✓ | | | | | | | | | | | | | | | |
| **iPTMnet** | 1 | | | ✓ | | | | | | | | | | | | | |
| **Jensen Lab Diseases** | 1 | | ✓ | | | | | | | | | | | | | | |
| **KEGG** | 4 | | ✓ | ✓ | ✓ | ✓ | | | | | | | | | | | |
| **KFPT** | 1 | | | | | | | | | ✓ | | | | | | | |

# KG SOURCES

| | Total # KG per source | Monarch KG | RTX-KG2 | ROBOKOP | Bioteque | SPOKE | GenomicKB | HetioNet | clinical KG | Petagraph | DrugMechDB | PrimeKG | NCATS Gard | HRA-KG | EmBiology | PharmKG | GNBR |
|---|---|---|---|---|---|---|---|---|---|---|---|---|---|---|---|---|---|
| **KinAce** | 1 | | | ✓ | | | | | | | | | | | | | |
| **LabeledIn** | 2 | | | | | ✓ | | ✓ | | | | | | | | | |
| **LIFEDB** | 1 | ✓ | | | | | | | | | | | | | | | |
| **LINCS 100** | 4 | | | ✓ | | ✓ | | ✓ | | ✓ | | | | | | | |
| **LungMAP Human Anatomy (LMHA)** | 1 | | | | | | | | | | | | | ✓ | | | |
| **Logical Observation Identifiers Names and Codes (LOINC)** | 1 | | ✓ | | | | | | | | | | | | | | |
| **Mammalian Phenotype Ontology** | 2 | ✓ | | | | | | | | | | | ✓ | | | | |
| **Mass Spectrometry Ontology** | 1 | | | | | | | | ✓ | | | | | | | | |
| **Mayo Clinic** | 1 | | | | | | | | | | | ✓ | | | | | |
| **MAXO** | 1 | ✓ | | | | | | | | | | | | | | | |
| **MEDI** | 2 | | | | | ✓ | | ✓ | | | | | | | | | |
| **Medication reference terminology (MED-RT)** | 1 | | ✓ | | | | | | | | | | | | | | |
| **MedGen** | 1 | | | | | | | | | | | | ✓ | | | | |
| **MedLine** | 2 | | | | | ✓ | | ✓ | | | | | | | | | |
| **MEDLINE Plus** | 1 | | ✓ | | | | | | | | | | | | | | |
| **MetabolomicsWorkbench** | 1 | | | ✓ | | | | | | | | | | | | | |
| **MetaCyc** | 1 | | | | | ✓ | | | | | | | | | | | |
| **Metathesaurus** | 1 | | ✓ | | | | | | | | | | | | | | |
| **MeSH** | 5 | | ✓ | | | ✓ | | ✓ | | | ✓ | | ✓ | | | | |
| **MGI** | 3 | ✓ | | ✓ | | | | | | ✓ | | | | | | | |

# KG SOURCES

| | Total # KG per source | Monarch KG | RTX-KG2 | ROBOKOP | Bioteque | SPOKE | GenomicKB | HetioNet | clinical KG | Petagraph | DrugMechDB | PrimeKG | NCATS Gard | HRA-KG | EmBiology | PharmKG | GNBR |
|---|---|---|---|---|---|---|---|---|---|---|---|---|---|---|---|---|---|
| miRBase | 1 | | ✓ | | | | | | | | | | | | | | |
| Molecular Interactions Controlled Vocabulary (MI) | 1 | | ✓ | | | | | | | | | | | | | | |
| Molecular Interactions Ontology | 1 | | | | | | | | ✓ | | | | | | | | |
| Molecular Signatures Database (MSigDB) | 1 | | | | | | | | | ✓ | | | | | | | |
| MonarchKG | 1 | | | ✓ | | | | | | | | | | | | | |
| MONDO | 5 | ✓ | ✓ | ✓ | | | | | | | | ✓ | ✓ | | | | |
| MotifMap | 1 | | | | | | ✓ | | | | | | | | | | |
| MPATH | 1 | ✓ | | | | | | | | | | | | | | | |
| Mouse gene-to-phenotype mappings (MPMGI) | 1 | | | | | | | | | ✓ | | | | | | | |
| MTBBASE | 1 | ✓ | | | | | | | | | | | | | | | |
| MutationDs | 1 | | | | | | | | ✓ | | | | | | | | |
| NCBI Gene | 2 | ✓ | ✓ | | | | | | | | | | | | | | |
| NCBI Taxonomy | 3 | | ✓ | | | ✓ | | | | | ✓ | | | | | | |
| NCBO BioPortal | 2 | | | | | | | | | ✓ | | | ✓ | | | | |
| National Cancer Institute (NCI) | 1 | | ✓ | | | | | | | | | | | | | | |
| National Drug data File | 1 | | ✓ | | | | | | | | | | | | | | |
| National drug data file-reference terminology | 1 | | ✓ | | | | | | | | | | | | | | |
| Neuro Behavior Ontology (NBO) | 2 | ✓ | ✓ | | | | | | | | | | | | | | |
| NTNU-SB | 1 | ✓ | | | | | | | | | | | | | | | |
| OBA | 1 | ✓ | | | | | | | | | | | | | | | |

# KG SOURCES

| | Total # KG per source | Monarch KG | RTX-KG2 | ROBOKOP | Bioteque | SPOKE | GenomicKB | HetioNet | clinical KG | Petagraph | DrugMechDB | PrimeKG | NCATS Gard | HRA-KG | EmBiology | PharmKG | GNBR |
|---|---|---|---|---|---|---|---|---|---|---|---|---|---|---|---|---|---|
| **OBO Foundry** | 2 | | ✓ | | | | | | | ✓ | | | | | | | |
| **Offsides** | 1 | | | | ✓ | | | | | | | | | | | | |
| **OMIM** | 6 | ✓ | ✓ | ✓ | | ✓ | | | | | | | ✓ | | | ✓ | |
| **Omnipath** | 1 | | | | ✓ | | | | | | | | | | | | |
| **OncoKB** | 1 | | | | | | | | ✓ | | | | | | | | |
| **Ontology of Genes and Genomes** | 1 | | | | | | | | | | | | ✓ | | | | |
| **Open Targets** | 1 | | | | ✓ | | | | | | | | | | | | |
| **Orphanet** | 5 | ✓ | ✓ | ✓ | | | | | | | | ✓ | ✓ | | | | |
| **PANTHER** | 2 | ✓ | | ✓ | | | | | | | | | | | | | |
| **Parkinsonsuk-UCL** | 1 | ✓ | | | | | | | | | | | | | | | |
| **PathBank** | 1 | | ✓ | | | | | | | | | | | | | | |
| **Pathway Interaction Database** | 2 | | | | | ✓ | | ✓ | | | | | | | | | |
| **PATRIC** | 1 | | | | | ✓ | | | | | | | | | | | |
| **PathoPhenoDB** | 1 | | | | | ✓ | | | | | | | | | | | |
| **PATO** | 1 | ✓ | | | | | | | | | | | | | | | |
| **PDSP** | 1 | | | ✓ | | | | | | | | | | | | | |
| **PharmacoDB** | 1 | | | | ✓ | | | | | | | | | | | | |
| **PharmGKB** | 1 | | | | | | | | | | | | | | | ✓ | |
| **Pharos** | 1 | | | ✓ | | | | | | | | | | | | | |
| **Phenio** | 1 | ✓ | | | | | | | | | | | | | | | |

# KG SOURCES

| | Total # KG per source | Monarch KG | RTX-KG2 | ROBOKOP | Bioteque | SPOKE | GenomicKB | HetioNet | clinical KG | Petagraph | DrugMechDB | PrimeKG | NCATS Gard | HRA-KG | EmBiology | PharmKG | GNBR |
|---|---|---|---|---|---|---|---|---|---|---|---|---|---|---|---|---|---|
| **Phenotype and Trait Ontology (PATO)** | 2 | | ✓ | | | | | | | | | | ✓ | | | | |
| **PHI-BASE** | 0 | | | | | | | | | | | | | | | | |
| **PhosphoSitePlus** | 2 | | | ✓ | | | | | ✓ | | | | | | | | |
| **Physician data query** | 1 | | ✓ | | | | | | | | | | | | | | |
| **PINC** | 1 | ✓ | | | | | | | | | | | | | | | |
| **Pfam** | 3 | | | | | ✓ | | | ✓ | | ✓ | | | | | | |
| **PomBase** | 1 | ✓ | | | | | | | | | | | | | | | |
| **PPI** | 1 | | | | | | | | | | | ✓ | | | | | |
| **PREDICT** | 2 | | | | | ✓ | | ✓ | | | | | | | | | |
| **PRISM** | 1 | | | | ✓ | | | | | | | | | | | | |
| **PROGENy** | 1 | | | | | | | | | | | | | | | ✓ | |
| **ProtCID** | 1 | | | | | ✓ | | | | | | | | | | | |
| **Protien Modification Ontology** | 2 | | ✓ | | | | | | ✓ | | | | | | | | |
| **Protein Ontology** | 3 | ✓ | | | | | | | | | ✓ | | ✓ | | | | |
| **Provisional Cell Ontology (PCL)** | 1 | | | | | | | | | | | | | ✓ | | | |
| **Psychological index terms** | 1 | | ✓ | | | | | | | | | | | | | | |
| **Reactome** | 8 | ✓ | ✓ | ✓ | ✓ | ✓ | | ✓ | | | ✓ | ✓ | | | | | |
| **Reaxy** | 1 | | | | | | | | | | | | | | ✓ | | |
| **RefSeq** | 2 | | | | | | | | ✓ | ✓ | | | | | | | |
| **RegNetwork** | 1 | | | | | | ✓ | | | | | | | | | | |

# KG SOURCES

| | Total # KG per source | Monarch KG | RTX-KG2 | ROBOKOP | Bioteque | SPOKE | GenomicKB | HetioNet | clinical KG | Petagraph | DrugMechDB | PrimeKG | NCATS Gard | HRA-KG | EmBiology | PharmKG | GNBR |
|---|---|---|---|---|---|---|---|---|---|---|---|---|---|---|---|---|---|
| **Relation Ontology (RO)** | 3 | ✓ | ✓ | | | | | | | | | | | ✓ | | | |
| **RepoDB** | 1 | | | | ✓ | | | | | | | | | | | | |
| **Repohub** | 1 | | | | ✓ | | | | | | | | | | | | |
| **RGD** | 2 | ✓ | | ✓ | | | | | | | | | | | | | |
| **RHEA** | 1 | ✓ | | | | | | | | | | | | | | | |
| **RNAcentral** | 2 | ✓ | | | | | ✓ | | | | | | | | | | |
| **ROSLIN-Institute** | 1 | ✓ | | | | | | | | | | | | | | | |
| **RxNorm** | 1 | | ✓ | | | | | | | | | | | | | | |
| **SCREEN** | 1 | | | | | | ✓ | | | | | | | | | | |
| **Sequence Ontology** | 1 | ✓ | | | | | | | | | | | | | | | |
| **SemMedDB** | 1 | | ✓ | | | | | | | | | | | | | | |
| **SGD** | 2 | ✓ | | ✓ | | | | | | | | | | | | | |
| **SIDER** | 7 | | | ✓ | ✓ | ✓ | | ✓ | ✓ | | | ✓ | | | | ✓ | |
| **SIGNOR** | 2 | | | ✓ | | | | | ✓ | | | | | | | | |
| **SMPDB** | 2 | | ✓ | | | | | | ✓ | | | | | | | | |
| **SNOMED-CT** | 1 | | | | | | | | ✓ | | | | | | | | |
| **STARGEO** | 2 | | | | | ✓ | | ✓ | | | | | | | | | |
| **STITCH** | 1 | | | | | | | | ✓ | | | | | | | | |
| **STRING** | 6 | ✓ | | ✓ | ✓ | ✓ | | | ✓ | ✓ | | | | | | | |
| **SYNGO** | 1 | ✓ | | | | | | | | | | | | | | | |

# KG SOURCES

| | Total # KG per source | Monarch KG | RTX-KG2 | ROBOKOP | Bioteque | SPOKE | GenomicKB | HetioNet | clinical KG | Petagraph | DrugMechDB | PrimeKG | NCATS Gard | HRA-KG | EmBiology | PharmKG | GNBR |
|---|---|---|---|---|---|---|---|---|---|---|---|---|---|---|---|---|---|
| SYNGO-UCL | 1 | ✓ | | | | | | | | | | | | | | | |
| SYSCILLIA-CCNET | 1 | ✓ | | | | | | | | | | | | | | | |
| Thesaurus | 1 | | | | | | | | | | | | ✓ | | | | |
| TISSUES | 5 | | | ✓ | ✓ | ✓ | | ✓ | ✓ | | | | | | | | |
| TTD | 1 | | | | | | | | | | | | | | | ✓ | |
| Uberon | 7 | ✓ | ✓ | | | ✓ | ✓ | ✓ | | | ✓ | | | ✓ | | | |
| UBKG | 1 | | | | | | | | | ✓ | | | | | | | |
| UniChem | 1 | | ✓ | | | | | | | | | | | | | | |
| Units Ontology | 1 | | | | | | | | ✓ | | | | | | | | |
| UniProt | 7 | ✓ | ✓ | ✓ | | ✓ | | | ✓ | ✓ | ✓ | | | | | | |
| UMLS | 3 | | ✓ | | | | | | | ✓ | | ✓ | | | | | |
| UPHENO | 1 | ✓ | | | | | | | | | | | | | | | |
| Vasculature Common Coordinate Framework (VCCF) | 1 | | | | | | | | | | | | | ✓ | | | |
| Veterans Affairs National Drug File (VANDF) | 2 | | ✓ | | | | | | | | | | ✓ | | | | |
| Wbbt | 1 | ✓ | | | | | | | | | | | | | | | |
| WBLS | 1 | ✓ | | | | | | | | | | | | | | | |
| WBPhenotype | 1 | ✓ | | | | | | | | | | | | | | | |
| WikiPathways | 2 | | | | | ✓ | | ✓ | | | | | | | | | |
| Wikipedia articles | 1 | | | | | | | | | | ✓ | | | | | | |
| WormBase | 1 | ✓ | | | | | | | | | | | | | | | |

# KG SOURCES

| | Total # KG per source | Monarch KG | RTX-KG2 | ROBOKOP | Bioteque | SPOKE | GenomicKB | HetioNet | clinical KG | Petagraph | DrugMechDB | PrimeKG | NCATS Gard | HRA-KG | EmBiology | PharmKG | GNBR |
|---|---|---|---|---|---|---|---|---|---|---|---|---|---|---|---|---|---|
| **XAO** | 1 | ✓ | | | | | | | | | | | | | | | |
| **XenBase** | 1 | ✓ | | | | | | | | | | | | | | | |
| **XPO** | 1 | ✓ | | | | | | | | | | | | | | | |
| **Yubio Lab** | 1 | ✓ | | | | | | | | | | | | | | | |
| **Zfin** | 2 | ✓ | | ✓ | | | | | | | | | | | | | |
| **Zebrafish Phenotype Ontology** | 1 | ✓ | | | | | | | | | | | | | | | |
| **Zebrafish Anatomy and development Ontologies** | 1 | ✓ | | | | | | | | | | | | | | | |
| **ZFS** | 1 | ✓ | | | | | | | | | | | | | | | |
| **4DNucleome** | 2 | | | | | | ✓ | | | ✓ | | | | | | | |
| **Other - NLP/text mining** | 4 | | | ✓ | | | | | | | ✓ | | | | ✓ | | ✓ |
| **Other** | 4 | | | | ✓ | | ✓ | | | | ✓ | | | | ✓ | | |